\documentclass[%
reprint,prl,
amsmath,amssymb,
aps,
]{revtex4-2}

\usepackage{xparse}
\usepackage{braket}

\usepackage{mathtools}
\usepackage{appendix}
\usepackage{graphicx}% Include figure files
\usepackage{dcolumn}% Align table columns on decimal point
\usepackage{bm}% bold math
\usepackage{hyperref}% add hypertext capabilities
\usepackage{xcolor}
\usepackage[normalem]{ulem}
\usepackage{cancel}

\renewcommand{\vec}[1]{\mathbf{#1}}

\usepackage{float}
\usepackage{braket}
\usepackage{bbold}
\usepackage{mathtools}

\DeclareMathOperator{\Tr}{Tr}

\newcommand{\vex}[1]{\bm{\mathrm{#1}}}
\newcommand{\msf}[1]{\mathsf{#1}}

\newcommand{\D}{\mathcal{D}}

\newcommand{\suml}[1]{\sum\limits_{#1}}

\newcommand{\ww}{\omega}

\newcommand{\e}{\varepsilon}

\newcommand{\rb}{\vex{r}}

\newcommand{\tel}{\tau_{\msf{el}}}

\newcommand{\bpsi}{\bar{\psi}}

\newcommand{\Qsp}{Q_{sp}}
\newcommand{\tQ}{\tilde{Q}}

\newcommand{\floor}[1]{\lfloor #1 \rfloor}

\begin{document}
	
	\title{Universal Dephasing Mechanism of Many-Body Quantum Chaos}

	\author{Yunxiang Liao}
	\author{Victor Galitski}
	\affiliation{Joint Quantum Institute and Condensed Matter Theory Center, Department of Physics, University of Maryland, College Park, MD 20742, USA.}
	
	\date{\today}
	
	\begin{abstract}
Ergodicity is a fundamental principle of statistical mechanics underlying the behavior of generic quantum many-body systems. 
However, how  this universal many-body quantum chaotic regime emerges due to interactions remains largely a puzzle. This paper
demonstrates using both heuristic  arguments and a microscopic calculation that a dephasing mechanism, similar to 
Altshuler-Aronov-Khmelnitskii dephasing in the theory of localization, underlies this transition to chaos. We focus on the behavior 
of the spectral form factor (SFF) as a function of  ``time'' $t$, which characterizes level correlations in the many-body spectrum.
The SFF can be expressed as a sum over periodic classical orbits and its behavior hinges on the interference of trajectories related to each other by a time translation. In the absence of interactions, time-translation symmetry is present for each individual particle, which leads to a fast exponential growth of the SFF and correspondingly loss of correlations between many-body levels. Interactions lead to dephasing, which disrupts interference, and breaks the massive time-translation symmetry down  to a global time-translation/energy conservation. This in turn gives rise to the hallmark linear-in-$t$ ramp in the SFF reflecting Wigner-Dyson level repulsion. This general picture is supported by a microscopic analysis of an interacting many-body model. Specifically, we study the complex $\mbox{SYK}_2+\mbox{SYK}_2^2$ model, which allows to tune between an integrable and chaotic regime. It is shown that the dephasing mass vanishes in the  former case, which maps to the non-interacting complex $\mbox{SYK}_2$ model via a time reparameterization. In contrast,  the chaotic regime gives rise to dephasing, which suppresses the exponential ramp of the non-interacting theory and  induces correlations between many-body levels.
	\end{abstract}

	\maketitle

There is a widely held quantum chaos conjecture~\cite{BGS,Prosen-chaos}, which states that the spectral statistics of quantum chaotic systems can be universally described by random matrix theory (RMT)~\cite{Wigner, Dyson, Mehta, Guhr,Kota}. 
This conjecture is supported by extensive experimental~\cite{WDexpt, WDexpt2, WDexpt3, WDexpt4, WDexpt5,WDexp6,WDexp7} and numerical studies~\cite{mcdonald,casati,Berry1981}, and 
has been used as one of the diagnostics of quantum chaos.
However, despite numerous efforts, the theoretical understanding of the underlying general mechanism behind this conjecture is far from complete. There have been several studies, which use the semiclassical periodic orbit theory~\cite{Gutzwiller} to prove that the spectral form factor -- a probe of two-level statistics -- indeed follows the RMT prediction for single-particle quantum chaotic systems~\cite{BerrySFF, Argaman, Sieber_2001, Sieber_2002, Muller_2004, Muller_2005}, including disordered metals~\cite{Argaman-disorder,Altshuer-PO}.
This approach has also been generalized to certain many-body systems with a well defined semiclassical limit~\cite{Muller-MB,Guhr-MB,Richter-MB,Urbina-MB}. In the context of the period-orbit theory,  the Wigner-Dyson level statistics stems from constructive interference between periodic paths related by a time translation. This time-translation argument has also been employed in studies of Floquet quantum circuits~\cite{ Chalker-1,Chalker-2,Chalker-3,Chalker-4,Chalker-5,Chalker-6,Prosen-chaos3}, periodically kicked interacting spin and fermionic chains~\cite{Prosen-chaos,Prosen-chaos2,Prosen-Fermion}, and the Sachdev-Ye-Kitaev (SYK) model~\cite{SSS}.
An alternative method widely used in the investigation of spectral statistics of chaotic systems is a field theoretical technique known as the non-linear $\sigma$ model, in supersymmetric~\cite{Efetov,Andreev-Ballistic, Andreev, Andreev2}, replica~\cite{Kamenev-GUE,Lerner,Kamenev} and Keldysh~\cite{Kamenev-Keldysh} frameworks.
This method allows the calculation of single-particle spectral correlations, but generalization to a microscopic analysis of many-body level statistics of interacting models~\cite{SYK-Altland,SYK-Altland-2,SYK-Altland-3} remains a challenge. 

In this Letter, we show that the emergence of many-body quantum chaos from the  non-interacting (single-particle chaotic) model can be generally understood in terms of dephasing of single-particle trajectories mediated by interactions.
This dephasing mechanism has some parallels with the Altshuler-Aronov-Khmelnitskii dephasing~\cite{AAK} in the weak localization (WL) theory of disordered metals~\cite{AA,AAG}, but with the characteristic energy level separation ($\sim t^{-1}$ in the SFF, $K(t)$) playing the role of the temperature $T$ in the context of WL. Specifically, we show that just like in the theory of WL, the behavior of the SFF is governed by a collective diffuson-like mode, which acquires an infrared cutoff in the presence of interactions as follows
\begin{align}
\label{dephasing}
\D_0(\Delta \tau) \xrightarrow{\mbox{interactions}} e^{-F_\phi(\Delta \tau)}\, \D_0(\Delta \tau).
\end{align}
Here $\D_0$ represents a gapless  collective mode in the non-interacting case, which in the Fourier space has the form $\D_0(\omega,{\bf q}) \propto \left( -i \omega + Dq^2\right)^{-1}$   in the case of a disordered metal ($D$ is the diffusion coefficient) and $\D_0(\omega) \propto 1/\omega$ in the RMT case of ``zero-dimensional diffuson.'' $\Delta \tau$, $\ww$, and $q$ are the time, frequency, and momentum coordinates, respectively.
The extra factor $e^{-F_\phi(\Delta \tau)}$ acquired by the interacting collective mode represents the exponential suppression in time stemming from the dephasing processes (the dephasing function $F_\phi(\Delta \tau \to \infty) = +\infty$).
The specific form of the dephasing function $F_{\phi}$ depends on the dimension and other details~\cite{dephasing-Pauli1,dephasing-Pauli2}.
%, but it generally is a power-law $\propto (t/\tau_\phi)^\alpha$ (e.g., for weak localization in d=2, it's simply $F_\phi(t) = t/\tau_\phi$, where $\tau_\phi$ is the AAK dephasing time). 

Below, we provide a brief qualitative review of the theory of WL and periodic-orbit theory of chaos and describe heuristic arguments underlying our main result. The second part of the paper is devoted to a microscopic $\sigma$-model analysis of an interacting many-body model, where we demonstrate the appearance of a dephasing cutoff of the relevant collective modes in the ergodic regime  and the absence thereof in the integrable limit. 

% Another technique is $\sigma$-model:...

%%%%%%%%%%%%%%%%%%%%%%%%%%%%%%%%%%%%%%%%%%%%%%%%%%%%%%
%	review dephasing for weak localization correction
%%%%%%%%%%%%%%%%%%%%%%%%%%%%%%%%%%%%%%%%%%%%%%%%%%%%%%

\begin{figure}[h!]
    \centering
	\includegraphics[width=0.8\linewidth]{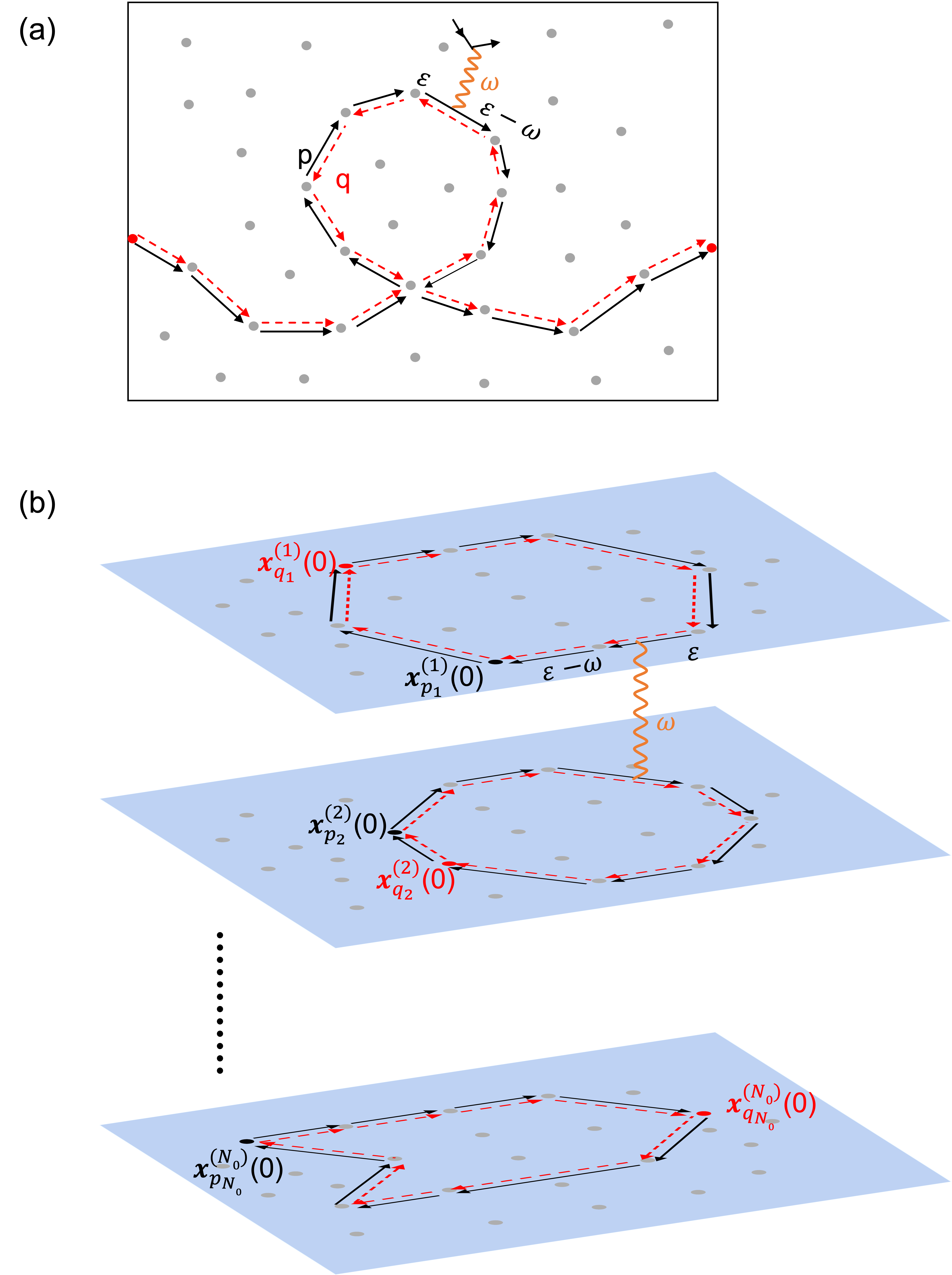}
	\caption{Qualitative picture of the dephasing mechanism for (a) weak localization and (b) periodic-orbit theory of chaos. 
	Panel (a) depicts two self-intersecting paths $p$ (black lines) and $q$ (red dashed line) of a particle moving in a disordered metal.
	The two paths coincide everywhere except for the loop which is traversed in the opposite directions by them. 
	The interference between such pair of paths leads to the WL correction to conductivity. In the presence of interactions, emissions of the electron-hole pairs (orange wavy line) result in the destruction of the phase coherence between the pair of paths and consequently a reduction in their quantum interference. 
	Panel (b) shows two periodic paths $\left\lbrace p\right\rbrace$ and $\left\lbrace q\right\rbrace$ of a many-body system in a chaotic medium.
	The $i$-th plane represents the phase space of the $i$-th particle, and the black solid and red dashed lines in that plane describe the corresponding particle's trajectories for periodic paths $p_i$ and $q_i$, respectively.
	Paths $\left\lbrace p\right\rbrace$ and $\left\lbrace q\right\rbrace$ are related by a ``individual time translation'' and differ only in the starting positions (denoted by black and red dots, respectively). 
	In the non-interacting case, the interference between such pair of paths is essential to the exponential-in-$t$ ramp in the SFF. In the presence of interactions, particle-particle collisions (orange wavy line) destroy the coherence unless the pair is related by a ``global time translation'', leading to the suppression of the exponential ramp. }
	\label{fig:p1}
\end{figure}

In a weakly disordered metal, the probability for a particle to diffuse from one point to another in the semiclassical limit can be expressed as~\cite{AA,AAG} 
\begin{align}\label{eq:P}
	{\cal P}=\left| \sum_{p} A_p e^{\frac{i}{\hbar}S_p}\right|^2=\sum_{pq}A_p A_q^*e^{\frac{i}{\hbar}S_p-\frac{i}{\hbar}S_q},
\end{align}
%The summation runs over all Feynman paths connecting these two points, $S_i$ and $A_i$ are, respectively, the action and amplitude associated with path $i$.
where the summation runs over all classical paths, labeled by $p$, connecting these two points. $S_p$ and $A_p$ represent, respectively, the action and amplitude of the path $p$.
In Eq.~\ref{eq:P}, the diagonal terms ($p=q$) contribute to the classical probability and are associated with the Drude conductivity, while the off-diagonal terms ($p \neq q$)  correspond to the quantum interference correction.
The interference between a generic pair of paths vanishes after disorder averaging due to the strong sensitivity of the action $S_p$ to the impurity potential.
However, there exist pairs of coherent paths, such as the self-intersecting ones in Fig.~\ref{fig:p1}(a), whose actions are identical in the absence of magnetic field, spin-orbit interactions, and particle-particle interactions.
Their interference can no longer be neglected, and results in the WL correction to the conductivity.
The WL correction to the conductivity is related to the probability of finding such self-intersecting paths~\cite{AA,AAG}
\begin{align}\label{eq:WL}
    \delta \sigma_{\rm WL} 
    \propto 
    -\int_{\tel}^{\infty} d\tau C(\rb,\rb;\tau)
    \propto
    -\int_{\tel}^{\infty} d\tau \frac{1}{(D\tau)^{d/2}}.
\end{align}
Here $C(\rb,\rb';\tau)$ represents the Cooperon (or the diffuson in the particle-particle channel), which is the Green function of the diffusion equation. Specifically,  $C(\rb,\rb;\tau)$ measures the return probability of a diffusing particle that starts at and returns to the same point $\rb$ in time $\tau$. $\tel$ indicates the elastic scattering time and $d$ is the dimension of the system. At this level, WL represents a single-particle effect and formally diverges in $d\leq 2$.
%in dimensions two and below. 

In the presence of interactions, the coherence between the time-reversed paths is reduced through the emissions of particle-hole pairs,
 and is completely destroyed when the traverse time exceeds the dephasing time $\tau_{\phi}$.
%%As an example, let us now consider the effect of one particle-hole emission to the pair of time-reversed paths depicted in Fig.~\ref{fig:p1}. 
%%Without interactions, their interference give rise to an additional $2|A_{p}|^2$ to the propagation probability. 
%For example, emission of one particle-hole pair of energy $\ww$ modifies the action of path $p/q$ in Fig.~\ref{fig:p1}(a) to $S_{p/q} \rightarrow S_{p/q}-\ww t^{em}_{p/q}$, where $t^{em}_{p/q}$ denotes the time it takes to traverse the trajectory after the emission.
%As a result, the interference between the pair is suppressed by a factor of $\cos \left( \ww(t^{em}_{p}-t^{em}_{q})\right) $.
It has been emphasized in Ref.~\cite{AAG,dephasing-Pauli1,dephasing-Pauli2} that the dephasing is dominated by real inelastic collisions with energy transfer $\tau_{\phi}^{-1} \ll \ww \ll T$.
Inelastic processes with energy transfer $\ww \gg T$ are not allowed since quasiparticles with energy $\e\sim T$ measured from the Fermi surface cannot lose energy $\ww \gg T$ due to the Pauli blocking.
When energy transfer $\ww \ll \tau_{\phi}^{-1}$, the action difference between the time-reversed pair can be ignored and therefore their coherence is preserved.
Dephasing processes result in the appearance of a mass term in the Cooperon ($\tau_{\phi}^{-1}$), which cuts off the WL integral (Eq.~\ref{eq:WL}) in the infrared limit.
%The exact calculation of dephasing rate for disordered metals at low temperature has been presented in Ref.~[AAK], see also Refs.~[...].

%\begin{widetext}

%\end{widetext}

To understand the effect of dephasing on spectral statistics, we now proceed with a sketch of the main idea of the periodic orbit derivation of the RMT spectral form factor (SFF) in the context of single-particle quantum chaos, see Refs.~\cite{Gutzwiller,Haake} for a more complete review.
The SFF measures the correlation between two energy levels and is defined as
\begin{align}\label{eq:K}
\begin{aligned}
	K(t)=\left\langle \Tr e^{-\frac{i}{\hbar}Ht} \Tr e^{\frac{i}{\hbar}Ht} \right\rangle,
\end{aligned}
\end{align}
where $H$ is the Hamiltonian of the system and the angular bracket stands for the ensemble averaging.
One can write the analytically continued partition function 
$\Tr e^{-\frac{i}{\hbar}Ht}=\int d^d{\vec{r}} \bra{\vec{r}} e^{-\frac{i}{\hbar}Ht} \ket{\vec{r}}$ as an integral of quantum propagation amplitude over the real space position ${\vec{r}}$.
The propagation amplitude, in the semiclassical limit, can be approximated as a summation over all classical paths that start and end at the same point ${\vec{r}}$ in time $t$.
Integration over ${\bf r}$ further restricts the summation to be over those paths whose initial and final momenta are also identical. 
As a result, only periodic paths that return to its starting point in the phase space in fixed time $t$ need to be retained. 
Grouping together all periodic paths that follow the same orbit and differ only in the starting points,
$\Tr e^{-\frac{i}{\hbar}Ht}$ can be represented by a sum over periodic orbits of period $t$~\cite{Gutzwiller}:
\begin{align}\label{eq:trU}
	\Tr e^{-\frac{i}{\hbar}Ht}
%	=\int dr \left\langle r| e^{-\frac{i}{\hbar}Ht} |r\right\rangle 
	=\sum_{P} A_P e^{\frac{i}{\hbar}S_P},
\end{align}
with $A_P$ and $S_P$ being the amplitude and action of the orbit $P$, respectively.
%\textcolor{blue}{(Here we have ignored a term which corresponds to 
%	$\left\langle \Tr e^{-\frac{i}{\hbar}Ht}\right\rangle $ and thus is not important to the connected SFF?)
%}
Here orbits that transverse the primitive orbit different numbers of times are considered as distinct, and the ones with multiple traversals will be ignored due to exponential proliferation of the primitive orbits~\cite{Hannay}.
We emphasize that each term in Eq.~\ref{eq:trU} represents the contribution from an infinite group of periodic paths that are related to each other by a time translation and share the same amplitude and action.
As a result, the orbit's amplitude $A_P$ contains a factor of $t$ originating from the integration over all possible time shifts.
Inserting Eq.~\ref{eq:trU} into the definition of the SFF (Eq.~\ref{eq:K}), one obtains
\begin{align}\label{eq:K-SC}
\begin{aligned}
	K(t)=\left\langle \sum_{P,Q} A_P A_Q^* e^{\frac{i}{\hbar}S_P-\frac{i}{\hbar}S_Q} \right\rangle .
\end{aligned}
\end{align}
%Note the close resemblance between this equation and Eq.~\ref{eq:P}. 
%However, the summation in Eq.~\ref{eq:P} runs over all paths that connect the given initial and final positions in the real space, whereas Eq.~\ref{eq:K-SC} sums over periodic orbits, each of which contains infinite periodic paths that return to their starting points in the phase space in time $t$.
In the absence of any symmetry, the off-diagonal terms ($P\neq Q$) in Eq.~\ref{eq:K-SC} vanishes upon ensemble averaging because of the highly oscillatory nature of the phases~\cite{BerrySFF}.
One can therefore keep only the diagonal terms ($P=Q$), whose total contribution can be obtained using the Hannay–Ozorido de Almeida sum rule~\cite{Hannay} $K(t)=\left\langle \sum_{P}|A_P|^2\right\rangle \simeq t/t_H$, valid for $t$ larger than the ergodic time but smaller than the Heisenberg time $t_H$.
This leads to the linear ramp of the SFF as predicted by RMT.
%In the presence of time-reversal symmetry, the interference between pairs of time-reserved orbits has also to be included, which results in an extra factor of $2$ in the SFF. 
%We note that nondiagonal contribution is important to recover the higher order in $t$ terms in the SFF.

%%%
We emphasize that, unlike the diagonal term in Eq.~\ref{eq:P} which is only responsible for the classical probability, the diagonal contribution to the SFF in Eq.~\ref{eq:K-SC} takes into account the quantum interference between periodic paths connected by a time translation. 
More specifically, for each periodic path $p$, one has to consider its interference with any path $q$ that is related to the original one by $\vex{x}_{q}(t')=\vex{x}_{p}(t'+\Delta)$ for arbitrary time shift $\Delta \in (0, t)$. Here $\vex{x}_p(t')$ denotes the phase-space coordinate at time $t'$ for the periodic path $p$ which obeys $\vex{x}_p(t)=\vex{x}_p(0)$. The factor of $t$ in the SFF originates from the integration over all possible time shifts $\int_0^{t} d\Delta=t$.
We can therefore see that the universal behavior of the spectral properties of quantum chaotic systems originates not from the specific details of the periodic orbits but from the quantum interference between periodic paths related by a time translation. 
For integrable systems,  Eq.~\ref{eq:K-SC} can still be used to express the SFF except now 
the family of periodic paths contributing to each term in Eq.~\ref{eq:trU} stay on a high dimensional torus instead of a closed curve in the chaotic case. 
These periodic paths are not necessary related by a time translation, and the above argument which relies on the integration over time shift is no longer applicable.

The semiclassical representation of the SFF Eq.~\ref{eq:K-SC} can also be generalized to many-body systems with well-defined semiclassical limit, and the summation in this case runs over periodic orbits (or tori for integrable cases) in the many-body phase space.
Before moving to many-body quantum chaos, let us now consider a system of non-interacting particles whose single-particle dynamics is chaotic. 
In the following, for an illustrative purpose, we ignore exchange statistics (indistinguishability of particles) and use a higher dimensional vector 
$\vex{X}_{\left\lbrace p\right\rbrace} (t')=\begin{bmatrix} \vex{x}^{(1)}_{p_1}(t'), \vex{x}^{(2)}_{p_2}(t'),..., \vex{x}^{(N_0)}_{p_{N_0}}(t')\end{bmatrix}$ 
to characterize the periodic path in the many-body phase space,
where $\left\lbrace p\right\rbrace$ represents the group of path indices for all particles. Its element is the individual particle coordinate $\vex{x}^{(i)}_{p_i}(t')$ in its own phase space at time $t'$, with $i$ being the particle index and $N_0$ the total number of particles.
The family of periodic paths belonging to the same periodic torus can be generated by individually time translate each single-particle path:
$\vex{x}^{(i)}_{q_i} (t')= \vex{x}^{(i)}_{p_i} (t'+\Delta^{(i)})$, where each time shift $\Delta^{(i)}$ can take different value within the regime $[0,t)$.
The special translation where all time shifts $\Delta^{(i)}$ are identical will be called ``global time translation''. 
%i.e. $\vex{X}_{\left\lbrace q \right\rbrace} (t')=\vex{X}_{\left\lbrace p\right\rbrace} (t'+\Delta)$ .
All these periodic paths connected by ``individual time translation'' share the same amplitude and action, and as a result, it is essential to take into account the quantum interference between all of them, see Fig.~1(b).
Counting the total number of periodic paths within the same torus, one would estimate that the SFF becomes a polynomial in $t$ at order $N_0$ instead.
This is consistent with the previous study, which evaluates the SFF of a system of non-interacting fermions populating the single-particle energy levels of a $N\times N$ random matrix from a Gaussian unitary ensemble (GUE), and finds a fast growing exponential-in-$t$ ramp~\cite{PRL,Sup}.
As explained in the Supplemental Material~\cite{Sup}, this exponential ramp is an approximation to a $\floor{N/2}$ order polynomial in the large $N$ limit.
Based on the periodic orbit discussion above, one can deduce that the contribution to the SFF is of the order of $t^{N_0}$ ($t^{N-N_0}$) when the number of fermions is $N_0\leq \floor{N/2}$ ($N_0>\floor{N/2}$). Taking into account all possible configurations, we arrive at a polynomial 
of the form $\sum_{n=1}^{\floor{N/2}} c_n t^{n}$,
%in $t$ at order $\floor{N/2}$, 
consistent with the analytical result~\cite{PRL,Sup}.
See also Refs.~\cite{Evans,Kostenberger} for a calculation of the SFF of an non-interacting Floquet model whose single-particle evolution over one period is governed by a Haar random unitary matrix.

Let us now introduce a generic type of interactions between the particles, which leads to a transition to many-body quantum chaos.
In this case, interactions destroy the phase coherence between the periodic paths connected by ``individual time translation'' but not those related by ``global time translation,'' which simply implies conservation of total energy.
The dephasing mechanism in this case is analogous to that in the theory of WL, and the coherence between the periodic paths connected by ``individual time translation'' is destroyed through the inelastic collisions between particles. 
However, the energy transfer $\ww$ of the dephasing processes are no longer required to be much smaller than $T$ (as one considers now the statistical properties of entire spectrum and the notion of temperature does not appear in the theory).
% \textcolor{blue}{ but should be much larger compared with the inverse of the period $t^{-1}$ (energy separation) to significantly modify the action.
% %The distribution function changes from $\cosh({\ww/T})-\tanh({\ww/T})$ for the weak localization case to the oscillation function $\cos({\ww t})-\tan({\ww t})$ for the SFF. 
% }
Since the number of coherent paths pairs is now significantly reduced due to the dephasing processes, the exponential ramp of the non-interacting model is suppressed to a linear one. 
As in the case of WL, the destruction of phase coherence is reflected by a mass acquired by the ``inter-replica'' diffuson -- the correlator of a particle and a hole governed by the forward and backward time-evolution operator $e^{-iHt}$ and $e^{iHt}$, respectively.
The presence of the mass in the ``inter-replica'' diffuson  suppresses its contribution to the SFF.
	
Focusing on the dephasing mechanism,
we compare the level statistics of an ensemble of chaotic systems with that of integrable systems, both of which are described by the following complex SYK$_2$ + SYK$_2^2$ Hamiltonian:
\begin{align}\label{eq:H}
\begin{aligned}
H=\sum_{i,j=1}^{N} \psi^{\dagger}_i h_{ij}  \psi_j 
+ 
\frac{U}{2} \left( \sum_{i,j=1}^{N}   \psi^{\dagger}_i V_{ij}  \psi_j\right)^2.
\end{aligned}
\end{align}
For the chaotic ensemble, $h$ and $V$ are $N\times N$ random Hermitian matrices drawn independently from the Gaussian Unitary Ensemble (GUE) with the same distribution function
\begin{align}\label{eq:Ph}
	P(M) \propto \exp \left( -\frac{N}{2J^2}\Tr M^2   \right), 
	\quad M=h, V.
\end{align}
For the integrable case, $h=V$ for each system in the ensemble and is a $N\times N$ GUE matrix that also follows the distribution function Eq.~\ref{eq:Ph}.
We consider weak enough interaction strength $U$ and perform a perturbative calculation in $U$.

In a previous study~\cite{PRL},  we investigated the SFF of the noninteracting model ($U=0$) and found an exponential-in-$t$ ramp. The eigenenergy $E_0$ of the noninteracting model  is simply related to the eigenenergy $E$ of the interacting integrable model (Eq.~\ref{eq:H} with $h=V$) by 
\begin{align}
	E^{(n)}=E_0^{(n)}+\frac{U}{2} (E_0^{(n)})^2,
\end{align}
where the superscript $n$ labels the energy level.
In the case of $h=V$, if one transforms to the basis where the single-particle Hamiltonian $h$ is diagonal, the many-body Hamiltonian then acquires the simple form
\begin{align}\label{eq:H2}
\begin{aligned}
	H=\sum_{i=1}^{N} \psi^{\dagger}_i \xi_{i}  \psi_i 
	+ 
	\frac{U}{2} \left( \sum_{i=1}^{N}   \psi^{\dagger}_i \xi_{i}  \psi_i\right)^2,
\end{aligned}
\end{align}
with $\xi_i$ being the $i$-th eigenvalue of the matrix $h$. One can see that the occupation number of each single-particle level is a conserved quantity, and the $h=V$ case is integrable.
The two-body interactions of the current model is of the SYK$_2^2$ form and different from that of a previous work by the authors~\cite{arXiv}. This explicit form is chosen so that the interactions can be Hubbard-Stratonovich decoupled via a  scalar field  (see details below), in an attempt to simplify the calculation. Moreover, it allows us to study in parallel the chaotic and integrable ensembles described by the same Hamiltonian Eq.~\ref{eq:H}.

We start from the following fermionic path integral representation of 
the SFF (Eq.~\ref{eq:K}) of the two ensembles described by Eq.~\ref{eq:H}:
\begin{align}\label{Seq:KT-1}
\begin{aligned}
&K(t)
=
\left\langle 
\int \D (\bpsi, \psi)  
\exp 
\left\lbrace 
i\sum_{a=\pm}
\int_0^{t^a} \!\!\! d t'
\left[ 
\bpsi_i^{a} (t') i \partial_{t'}  \psi_i^{a}(t')
\right. \right. \right. 
\\
&
\left. \left. \left. 
-\zeta_a
\bpsi_i^{a} (t') h_{ij} \psi_j^{a}(t')
-
\zeta_a\frac{U}{2} 
\left( \bpsi_i^{a} (t') V_{ij}\psi_j^{a} (t') \right)^2 
\right] 
\right\rbrace 
\right\rangle.
\end{aligned}	
\end{align}
Here the fermionic field $\psi^a_i$ carries a replica index $a=\pm$ (labels the forward/backward evolution) and a flavor index $i =
1,2,...,N$.
It is subject to the antiperiodic boundary condition $\psi^{a}(t^a)=-\psi^{a}(0)$. $\zeta_a=\pm 1$ for $a=\pm$, and
$t^a$ is defined as $t^{a}=t \mp i \zeta_a 0^+$, where the infinitesimal imaginary increment $\mp i \zeta_a 0^+$ plays a key role selecting the appropriate saddle points via spontaneous breaking of unitarity~\cite{arXiv}. To focus on the dephasing mechanism, the infinitesimal imaginary increment in $t^{a}$ is ignored in the following.
Through a standard derivation~\cite{Sup}, we obtain a $\sigma-$model representation of the SFF: %of these two ensembles  :
	\begin{align}\label{eq:sigma}
	\begin{aligned}
	&K(t)=\,
	\frac{1}{Z}
	\int \D\phi e^{-S_{0}[\phi]}
	\int \D Q \exp \left( - S[Q,\phi]\right), 
	\\
	&S_{0}[\phi]
	=
	-\sum_{a=\pm} \frac{i\zeta_a }{2U}  \int_0^{t} d t'
	\left( \phi^a_{t'} \right)^2,
	\\
	&S[Q,\phi]
	=\,	
	\frac{N}{2J^2}
	\Tr
	\left(QFQ \right)
	-
	N
	\Tr 
	\ln
	\left(
	i\sigma^3  \partial_{t}
	+
	i F Q
	\right).
	\end{aligned}
	\end{align}
	Here, $\sigma^3$ denotes the third Pauli matrix in the replica space, $Z$ is the normalization constant~\cite{Sup}.
	%and $\zeta_a=\pm 1$  for replica index $a=\pm$.
	The real bosonic field $\phi$ is introduced to decouple the interactions and is subject to the periodic boundary condition $\phi^a_{t}=\phi^a_{0}$,
	while the Hermitian matrix field $Q$ decouples the ensemble-averaging generated interactions and satisfies $Q^{ab}_{t_1+t,t_2}=-Q^{ab}_{t_1,t_2}$.
    $F$ takes different forms for the chaotic and integrable cases:
	\begin{align}\label{eq:F}
	\!\!\!F^{ba,a'b'}_{t_2 t_1, t_1' t_2'}
	=
	\begin{cases}
	\left(1+\phi^a_{t_1}\phi^b_{t_2}\right)
	{\bf 1}^{ba,a'b'}_{t_2 t_1, t_1' t_2'},
	&
	h\neq V,
	\\
	\left( 1+\phi^a_{t_1}\right)
	\left(1+\phi^b_{t_2}\right)
	{\bf 1}^{ba,a'b'}_{t_2 t_1, t_1' t_2'},
	&
	h=V.
	\end{cases}
	\end{align}	
	We emphasize that the seemingly small difference in $F$ between the chaotic and integrable models is responsible for the strikingly contrasting behaviors of their SFFs.

%	We have defined $t^{a}=t \mp i \zeta_a 0^+$, whose infinitesimal imaginary increment plays a key role selecting the appropriate saddle points via spontaneous breaking of unitary~\cite{arXiv}. However, to focus on the dephasing mechanism, the infinitesimal imaginary increment in $t^{a}$ is ignored in the following for simplicity.
	
In the integrable $h=V$ case, one can perform a time reparametrization: 
\begin{align}\label{eq:TR}
\begin{aligned}
	  Q^{ab}_{t_1t_2} \rightarrow Q^{ab}_{\tau^a(t_1) ,\tau^b(t_2) },
	\,\,
	\tau^a(t_1) \equiv \int_{0}^{t_1} d t' \left( 1+\phi^a_{t'} \right),
\end{aligned}
\end{align}
after which the action becomes that of the non-interacting theory $S[Q,\phi=0]$ with shifted time $t \rightarrow t+\bar{\phi}^{\pm}$. Here $\bar{\phi}^{\pm}$ is defined as $\bar{\phi}^{\pm}\equiv \int_0^{t} d t' \phi^{\pm}_{t'}$. The SFF of the integrable model is now related to that of the non-interacting model $H_0$ ($U=0$) by
\begin{align}
\begin{aligned}
	K(t)=
	\dfrac{	\int \D \phi e^{-S_0[\phi] }
		\left\langle 
		\Tr e^{ -i H_0 \left( t+\bar{\phi}^+\right) }
		\Tr e^{+ iH_0 \left( t+\bar{\phi}^-\right) }
		\right\rangle}{\int \D \phi e^{-S_0[\phi]}},
%	\\
%	S_0=-\sum_{a} \frac{i\zeta_a }{2U t^a} (\phi_0^a)^2
\end{aligned}
\end{align}
 and can be solved using the same approach employed in Ref.~\cite{PRL}.
 
 \begin{figure}[t!]
	\centering
	\includegraphics[width=0.9\linewidth]{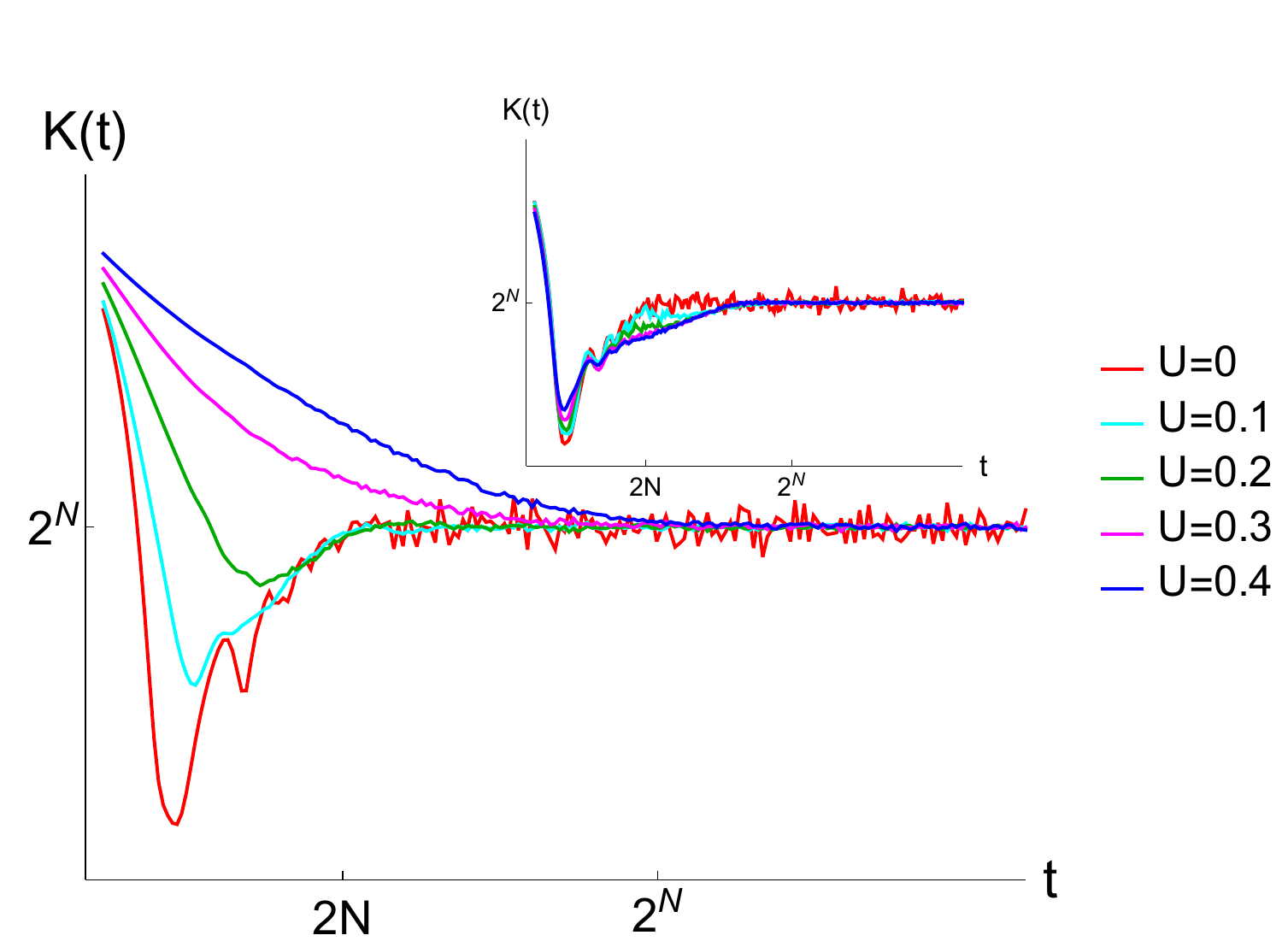}
	\caption{
	Numerical results of the SFF $K(t)$ for the complex SYK$_2$ + SYK$_2^2$ model (Eq.~\ref{eq:H}) in the integrable $h=V$ case and chaotic $h\neq V$ case (inset),
	plotted in the log-log scale.
	In both the main panel and the inset, the noninteracting SFF (red curve) is plotted for comparison.
	We set $N=10$ and $J=1$, and average over $1000$ samples.
% For small interaction strength $U$, the SFF of the integrable ensemble exhibits an exponential ramp  and a plateau that starts at $t \approx 2N$, as in the noninteracting case (red curve). For larger $U$, the ramp disappears and the slope reaches the plateau at a later time due to the nonuniversal contribution from the disconnected SFF.
% By contrast, the SFF of the chaotic ensemble exhibits a ramp which grows much slower compared with the noninteracting SFF.
	 }
	\label{fig:p2}
\end{figure}

In Fig.~\ref{fig:p2}, we show the numerical results of the SFFs in the integrable $h=V$ case for various interaction strengths $U$. For small interacting strength $U$, the SFF behaves similarly as its noninteracting counterpart (red curve), and exhibits a slope, an exponential ramp as well as a plateau that starts at time $t_p \sim N$.
At larger $U$, the exponential ramp disappears and the SFF exhibits a nonuniversal slope approaching the plateau at a later time.
This late plateau time is due to the fact that the nonuniversal slope contains a significant contribution from the slowly decaying disconnected SFF.
The SFF for the $h\neq V$ chaotic case is plotted in the inset of Fig.~\ref{fig:p2}. It exhibits a ramp which grows much slower and reaches the plateau at a much later time.
We note that, to see the linear in $t$ dependence of the ramp, a computation of a larger system size is required, see for example Refs.~\cite{SYK-Lau,SYK-Sun} for numerical works studying the level statistics of similar models.

We emphasize that, for both the chaotic and integrable ensembles, the action (Eq.~\ref{eq:sigma}) remains invariant under the aforementioned ``global time translation'':
\begin{align}
	\phi^{a}_{t_1} \rightarrow \phi^{a}_{t_1+\Delta^a}, 
	\qquad
	Q^{ab}_{t_1, t_2} \rightarrow Q^{ab}_{t_1+\Delta^a, t_2+\Delta^b},
\end{align}
where $\Delta^a$ can take different values for different replica index $a$.
%\begin{align}
%	&
%	\phi^{+}(t) \rightarrow \phi^{+} (t+\Delta), 
%	\qquad 
%	\phi^{-}(t) \rightarrow \phi^{-} (t)
%	\\
%	&
%	\begin{bmatrix}
%	Q^{++}_{t_1, t_2} & Q^{+-}_{t_1, t_2}
%	\\
%	Q^{-+}_{t_1, t_2} & Q^{--}_{t_1, t_2}
%	\end{bmatrix}
%	\rightarrow
%	\begin{bmatrix}
%	Q^{++}_{t_1+\Delta, t_2+\Delta} & Q^{+-}_{t_1+\Delta, t_2}
%	\\
%	Q^{-+}_{t_1, t_2+\Delta} & Q^{--}_{t_1, t_2}
%	\end{bmatrix}.
%\end{align}
For the integrable case, the theory possesses a larger symmetry group.
Under any $U(2)$ rotation applying to %the rescaled $Q$ matrix subblock
$
 	\begin{bmatrix}
 	\bar{Q}^{++}_{nn}/(t+\bar{\phi}^+)
 	& 
 	\bar{Q}^{+-}_{nm}/\sqrt{(t+\bar{\phi}^+)(t+\bar{\phi}^-)}
 	\\
 	\bar{Q}^{-+}_{mn}/\sqrt{(t+\bar{\phi}^+)(t+\bar{\phi}^-)}
 	& \bar{Q}^{--}_{mm}/(t+\bar{\phi}^-)
	\end{bmatrix},
$
the reparametrized action is invariant for any pair of energy indices $n,m$ satisfying 
$\frac{n+1/2}{t+\bar{\phi}^+}=-\frac{m+1/2}{t+\bar{\phi}^-}$.

We next perform a perturbative calculation for both the chaotic and integrable models. We consider first the integration over the matrix field $Q$, in the presence of arbitrary $\phi$.
In this case, the problem becomes similar to that of a quantum dot in the presence of a time-dependent perturbation~\cite{Kravtsov-DL,Skvortsov-DL,Aleiner-DL1,Aleiner-DL2}.
Because of the large overall factor $N$ in the action $S[Q,\phi]$, we approximate the integral with the contribution from saddle points $\Qsp$ and the Gaussian fluctuations $\delta Q$ around them.
Taking variation of the action with respect to $Q$, we find the saddle point equation 
%in the presence of arbitrary $\phi$:
\begin{align}	
	Q_{sp}
	=
	J^2
	\left(
	\sigma^3 \partial_{t}
	+
	F Q_{sp}
	\right)^{-1}.
\end{align}
The explicit expression for its solution up to order $\phi^2$ is provided in the Supplemental Material (Eqs.~(S38) and ~(S39))~\cite{Sup}. At $\phi^0$ order, the solution is the non-interacting saddle point, which in the Matsubara frequency space is given by
\begin{align}
\begin{aligned}
(Q_{sp}^{(0)})_{nn'}^{aa'}
=
\delta_{nn'}\delta_{aa'}
\frac{t}{2}
\left( 
i\zeta_a\e_n
+
s_n^a
\sqrt{4J^2 -(\e_n)^2}
\right).
%\\
%\approx&
%J s_n^a t^a \delta_{nn'}\delta_{aa'},
\end{aligned}
\end{align}
Here $\e_n=2\pi (n+1/2)/t$ is the fermionic Matsubara frequency. $s_n^a$ can take values of $+1$ and $-1$ when $|\e_n|\leq 2J$, and for $|\e_n|>2J$ is determined by the fact that $(\Qsp^{(0)})^{aa}_{nn} \rightarrow 0$ in the limit $ J\ll |\e_n|$~\cite{Kamenev-Keldysh,Winer}.
There are various saddle points corresponding to different choices of $\left\lbrace s_n^a\right\rbrace$~\cite{arXiv}.
Note that here we obtain an interacting saddle point $\Qsp$ in the presence of arbitrary configuration of $\phi$, in
contrast to the non-interacting saddle (with $\phi=0$) usually considered in the conventional Finkel'stein NL$\sigma$M calculation (e.g., Refs.~\cite{FNLsM,arXiv}).

The action for the Gaussian fluctuations around any saddle point $\Qsp$ can be expressed as 
\begin{align}\label{eq:SQ2}
\begin{aligned}
%&S[\tilde{Q}_{sp}+\delta\tilde{Q},\phi]-S[\tilde{Q}_{sp},\phi]
\delta S[\delta\tilde{Q},\phi]
=
\frac{N}{2}
\Tr
\left(
\delta \tQ
%^{a',b'}_{t_1',t_2'} 
M
%^{b',a';a,b}_{t_2',t_1';t_1,t_2}
\delta \tQ
%^{b,a}_{t_2,t_1},
\right),
%\\
%=\,&
%S_{sp}
%+
%\frac{N}{2}
%\sum \int
%\delta Q^{a',b'}_{t_1',t_2'} 
%F^{b',a';a',b'}_{t_2',t_1';t_1',t_2'}
%M^{b',a';a,b}_{t_2',t_1';t_1,t_2}
%F^{b,a;a,b}_{t_2,t_1;t_1,t_2}
%\delta Q^{b,a}_{t_2,t_1}
\end{aligned}
\end{align}
where $\delta \tilde{Q}=F\delta Q$~\cite{Fnb}. See Eqs.~(S58) and ~(S59) in the Supplemental Material~\cite{Sup} for the explicit expressions for $M$ for various saddle points.
The inverse of the kernel $M$ is related to the propagator of the Gaussian fluctuation:
\begin{align}\label{eq:D-a}
	\D^{ba,a'b'}_{t_2 t_1, t_1' t_2'}
	\equiv 
	N\left\langle \delta \tilde{Q}^{ba}_{t_2t_1} \delta \tilde{Q}^{a'b'}_{t_1't_2'} \right\rangle
	=
	(M^{-1})^{ba,a'b'}_{t_2 t_1, t_1' t_2'}.
\end{align}
The corresponding contribution from the Gaussian fluctuations around ${Q}_{sp}$ to the SFF is proportional to
\begin{align}\label{Seq:IntdQ}
\begin{aligned}
\int \D \delta \tilde{Q}
\exp \left( 
-\frac{N}{2}
\int
\delta \tQ
M
\delta \tQ
\right) 
\propto
%\exp \left( -\Tr \ln M  \right)
%=
\exp \left( \Tr \ln \D \right).
\end{aligned}
\end{align}

In the following, we will focus on the inter-replica fluctuations ($\delta Q^{ab} $ with $a\neq b$)  around a special saddle point $\Qsp^{(\pm)}$ with $s_n^a=\pm \zeta_a$ (for $|\e_n|\leq 2J$), whose kernel $M$ assumes a simple form~\cite{Sup}.
The fluctuations around other saddle points can be examined in an analogous way (although not as technically straightforward).
In the non-interacting theory, the inter-replica fluctuations around $\Qsp^{(\pm)}$ are massless, while
the intra-replica fluctuations, which contribute only to the nonuniversal disconnected SFF, are massive.
In the presence of interactions, for the integrable case, the inter-replica fluctuation propagator ($0$-dim diffuson) follows an equation equivalent to that of the non-interacting theory:
\begin{align}\label{eq:D-a}
\begin{aligned}
 \left[  
 \frac{\mp1}{2J^3} \left( \partial_{t_1}-\partial_{t_2}\right) 
\right] 
\mathcal{D}^{ba,a'b'}_{t_2 t_1, t_1' t_2'}
=
{\bf 1}^{ba,a'b'}_{t_2 t_1, t_1' t_2'},
\end{aligned}
\end{align}
consistent with results from time-reparametrization~\cite{Sup}.
By contrast, for the chaotic case, the diffuson equation becomes
\begin{align}\label{eq:D-b}
\begin{aligned}
	\left[  
	 \frac{\mp1}{2J^3}  \left( \partial_{t_1}- \partial_{t_2} \right) 
	+
	\frac{1}{2J^2}
	\left( \phi^{a}_{t_1}-\phi^{b}_{t_2} \right)^2 
	\right]  
	\mathcal{D}^{ba,a'b'}_{t_2 t_1, t_1' t_2'}
	=
	{\bf 1}^{ba,a'b'}_{t_2 t_1, t_1' t_2'}.
\end{aligned}
\end{align}
We emphasize the close resemblance of Eq.~\ref{eq:D-b} to the Cooperon equations for the interaction-induced dephasing of WL in disordered systems~\cite{AAK,AAG,dephasing,Davis,Schmid} and perturbation-induced dephasing of dynamical localization in driven quantum dots~\cite{Skvortsov-DL,Kravtsov-DL}.

While the inter-replica fluctuation remains massless for the integrable case, it acquires a mass arising from the $\phi$-dependent term in Eq.~\ref{eq:D-b} for the chaotic case.
For simplicity, let us now consider the large $t$ limit. The solution to Eq.~\ref{eq:D-b} can now be expressed as
\begin{align}\label{eq:D}
\begin{aligned}
    &\!\!\!\!
	\D^{ba,a'b'}_{t_2t_1,t_1't_2'}
	=
	\exp\left(
	-S_D[\phi]
	\right) 
	(\D_0)^{ba,a'b'}_{t_2t_1,t_1't_2'},
	\\
	&\!\!\!\!
	S_D[\phi]=\mp \frac{J}{2}
	\int_{\tau'}^{\tau} d v
	\left(
	\phi^a({u+v/2})-\phi^b({u-v/2})
	\right)^2 ,
	\\
	&\!\!\!\!
	(\D_0)^{ba,a'b'}_{t_2t_1,t_1't_2'}
    =
    \mp s' J^3\Theta\left(s'(\tau - \tau')\right) \delta(u-u')\delta_{aa'}\delta_{bb'},
\end{aligned}
\end{align}
where $t_{1,2}=u \pm \tau/2$ and $t_{1,2}'=u' \pm \tau'/2$.
$\D_0$ represents the solution to the non-interacting diffuson equation (Eq.~\ref{eq:D-a}), and $s'=\mp 1$ is determined by the boundary condition that $\D$ is non-divergent when $|\tau- \tau'|\rightarrow \infty$. 
We can therefore see that, in contrast to the integrable case, the Gaussian fluctuation propagator $\D$ in the chaotic case acquires an exponential factor $\exp\left(-S_D[\phi]\right)$, which decays with increasing $|\tau -\tau'|$ for almost all possible configurations of $\phi$ (except when $\phi^{a}_{t'}$ is time and replica index independent).

Substituting the Fourier transform of $\phi$ in $S_D[\phi]$ (Eq.~\ref{eq:D}), we find
\begin{align}\label{eq:DffusonEQ4}
\begin{aligned}
&S_D[\phi]
=
\mp 
\frac{J}{2t^2}
\suml{m}
\left(
|\phi^a_m|^2
+
|\phi^b_m|^2
-
2\phi^a_m \phi^b_{m}
e^{i2\ww_{m}u}
\right)
(\tau-\tau')
\\
&
\mp \frac{J}{t^2}
\suml{m\neq m'}\!\!\!
\frac{1}{i\ww_{m-m'}}
\left(
e^{i\ww_{m-m'}u}
\phi^a_m \phi^a_{-m'}
+
e^{-i\ww_{m-m'}u}
\phi^b_{-m} \phi^b_{m'}
\right. 
\\
&
\left. 
-
e^{i\ww_{m+m'}u}
2\phi^a_m \phi^b_{m'}
\right)
\left( e^{i\ww_{m-m'}\tau/2}-e^{i\ww_{m-m'}\tau'/2}\right). 
\end{aligned}
\end{align}
At large $\Delta \tau = |\tau-\tau'|$ and $|u|$, the dominant contribution to $S_D$ comes from the non-oscillating term $S_D[\phi]\approx |\tau - \tau'|/\tau_{\phi}$, which therefore determines the dephasing function $F_\phi[\Delta \tau] \approx \Delta\tau /\tau_\phi$ from Eq.~(\ref{dephasing}). The dephasing rate is as follows 
\begin{align}
    \tau_{\phi}^{-1}
    =
    \frac{J}{2t^2} 
    \left[
    \sum_{m\neq 0}(|\phi^{+}_m|^2+|\phi^{-}_m|^2) +(\phi^+_0-\phi^-_0)^2
    \right].
\end{align}
%\textcolor{blue}{ which is obtained by and keeping only the non-oscillating terms there (see Eq.~(S69) in the supplementary material).}

From Eq.~\ref{Seq:IntdQ}, we can see that the diffuson's contribution to the SFF is suppressed due to the extra exponential decaying factor $e^{-S_D[\phi]}$ acquired by the interacting diffuson in the chaotic case.
This decaying factor is a manifestation of the dephasing effect due to interactions.
The suppression of the exponential ramp is a necessary prerequisite for the expected transition from Poisson to RMT statistics, indicating a significant role played by the dephasing in the emergence of many-body quantum chaos.
The detailed calculation of the SFF for the chaotic model would require 
averaging $\det \D$ over the fluctuation of decoupling field $\phi$, where $\det \D$ arises from integration over the Gaussian fluctuations (Eq.~\ref{Seq:IntdQ}).
This is out of the scope of current study and is left to future work.

This work was supported by the U.S. Department of
Energy, Office of Science, Basic Energy Sciences under
Award No. DE-SC0001911. Y. L. acknowledges a post-
doctoral fellowship from the Simons Foundation ``Ultra-
Quantum Matter'' Research Collaboration. This work was performed at the Aspen Center for Physics, which is supported by National Science Foundation grant PHY-1607611 and Simons Foundation.

\bibliography{main}
		
\end{document}

\title{
		Universal Dephasing Mechanism of Many-Body Quantum Chaos
		\\
		Supplemental Material
	}
	\author{Yunxiang Liao}
	\author{Victor Galitski}
	\affiliation{Joint Quantum Institute and Condensed Matter Theory Center, Department of Physics, University of Maryland, College Park, MD 20742, USA.}
	
	\date{\today}

	\maketitle
	
	\tableofcontents

	\section{Introduction}
	
	In this supplementary material, we provide a detailed analysis of the spectral form factor (SFF) of two different ensembles described by the Hamiltonian:
	\begin{align}\label{Seq:H}
	\begin{aligned}
	H=\sum_{i,j=1}^{N} \psi^{\dagger}_i h_{ij}  \psi_j 
	+ 
	\frac{U}{2} \left(  \sum_{i,j=1}^{N}  \psi^{\dagger}_i V_{ij}  \psi_j\right)^2 .
	\end{aligned}
	\end{align}
	Specifically, we consider the following two cases:
	\begin{enumerate}
		\item  Chaotic ensemble ($h\neq V$)
		
		In this case, $h$ and $V$ are two statistically independent $N\times N$ random  Gaussian unitary ensemble (GUE) matrices with the same %$\left\langle h_{ij} \right\rangle =\left\langle V_{ij} \right\rangle $ and
		 two-point correlation $\left\langle h_{ij} h_{j'i'}\right\rangle=\left\langle V_{ij} V_{j'i'}\right\rangle=J^2/N\delta_{ii'}\delta_{jj'}$. Note that the randomness of the interaction strength can be independently controlled by the parameter, $U$. Since $h$ and $V$ are independent, their correlator $\left\langle h_{ij} V_{j'i'}\right\rangle=\left\langle h_{ij}\right\rangle\left\langle  V_{j'i'}\right\rangle=0$.
		The SFF of this ensemble can be written as:
		\begin{align}
		\begin{aligned}
		K(t)
		=
		\frac{1}{Z_h^2}
		\int \D h \D V
		e^{ -\frac{N}{2J^2}\Tr h^2 } e^{ -\frac{N}{2J^2}\Tr V^2 } 
		\Tr e^{ -iH t } \Tr e^{ +iHt  },
		\end{aligned}
		\end{align}
		where $Z_h=	\int \D h e^{ -\frac{N}{2J^2}\Tr h^2 }$ is the normalization constant.
		
		\item  Integrable ensemble ($h=V$)
		
		For the integrable case, $h$ is drawn from the same GUE as in the previous case. However, $h$ and $V$ are no longer independent but are equivalent for each system in this ensemble.
		Therefore, in this case, we have
		\begin{align}
		\begin{aligned}
		K(t)
		=
		\frac{1}{Z_h}
		\int \D h 
		e^{ -\frac{N}{2J^2}\Tr h^2 } 
		\left( \Tr e^{ -iH t } \Tr e^{ +iHt  } \right) \bigg\lvert_{h=V}.
		\end{aligned}
		\end{align}
		
	\end{enumerate}
	
	In the following, we will study the SFFs of these two different ensembles, and show that dephasing is the underlying mechanism that is responsible for the difference in the spectral statistics of chaotic and integrable systems. 
	
	\section{Derivation of the $\sigma$-model}
	
	The SFF of the ensembles described by Eq.~\ref{Seq:H} can be expressed as the following path integral:
		\begin{align}\label{Seq:KT-1}
		\begin{aligned}
		K(t)
		=&
		\int \D (\bpsi, \psi)  
		\left\langle 
		\exp 
		\left\lbrace 
		\begin{aligned}
		&i\sum_{a=\pm}
		\int_0^{t^a} d t'
		\left[ 
		\bpsi_i^{a} (t') (i \partial_{t'} \delta_{ij}-\zeta_a h_{ij}) \psi_j^{a}(t')
		-
		\zeta_a\frac{U}{2} 
		\left( \bpsi_i^{a} (t') V_{ij}\psi_j^{a} (t') \right)^2 
		\right] 
		\end{aligned}	
		\right\rbrace 
		\right\rangle.
		\end{aligned}	
		\end{align}
	Here, the Grassmann field $\psi$ is subject to the anti-periodic boundary condition $\psi^{a}(t^a)=-\psi^{a}(0)$,
	and the integrations over $\psi^{+}$ and $\psi^{-}$ give rise to $\Tr e^{-iHt}$ and $\Tr e^{+iHt}$, respectively. 
	In additional to the replica index $a$, $\psi^a_i$ also carries a flavor index $i=1,2,...,N$.
	The angular bracket represents ensemble averaging, and  $\zeta_a=\pm 1$ corresponds to the replica index $a=\pm$ (i.e., the forward/backward evolution).
	%	We have employed the convention that repeated indices implies summation.
	
	We note that the upper limit of the time integration
	in Eq.~\ref{Seq:KT-1} is defined as $t^{a}=t\mp i \zeta_a 0^+$ and the SFF described by this equation is in fact 
	$K(t)=\left\langle e^{-iH (t\mp i 0^+)}\Tr e^{iH (t\pm i 0^+)}\right\rangle$. It has been shown in the previous study~\cite{arXiv} that the infinitesimal imaginary increment in $t^a$ is needed to select the appropriate saddle points via spontaneous breaking of unitary. However, for the sake of simplicity, this infinitesimal imaginary increment will be ignored as our focus is on the dephasing mechanism.

	Decoupling the interaction term by Hubbard-Stratonovich (H.S.) transformation, one arrives at
		\begin{align}\label{Seq:K-1}
		\begin{aligned}
		&K(t)
		=
		\frac{1}{Z_{\phi}}\int D\phi e^{-S_0[\phi]}
		\int \D (\bpsi, \psi)  
		\left\langle 
		\exp 
		\left\lbrace 
		\begin{aligned}
		&i\sum_{a=\pm}
		\int_0^{t} d t'\,
		\bpsi_i^{a} (t') 	
		\left[ i \partial_{t'} \delta_{ij}-\zeta_a h_{ij}-\zeta_a \phi^a(t')V_{ij}\right]  
		\psi_j^{a}(t')
		\end{aligned}	
		\right\rbrace 
		\right\rangle ,
		\\
		&
		S_0[\phi]
		=
		-\sum_{a=\pm} \frac{i\zeta_a }{2U}  \int_0^{t} d t'
		\left( \phi^a(t')  \right)^2,
		\\
		&Z_{\phi}=\int D\phi e^{-S_0[\phi]}.
		\end{aligned}	
		\end{align}
		Here the bosonic H.S. field $\phi$ satisfies the periodic boundary condition: $\phi^a(t)=\phi^a(0)$.
		Effectively, the SFF can be expressed as
		\begin{align}
		\begin{aligned}
		&K(t)=
		\frac{1}{Z_{\phi}}
		\int D\phi
		e^{-S_0[\phi]}
		\left\langle 
		\Tr \exp\left[ -i\int_{0}^{t} dt' H^+_{\phi}(t')\right] 
		\Tr \exp \left[ i\int_{0}^{t} dt' H^-_{\phi}(t') \right] 
		\right\rangle,
		\\
		&H^{a}_{\phi}(t')=	
		\sum_{i,j=1}^{N}
		\psi^{\dagger}_{i}
		\left( h_{ij}+ \phi^a(t')V_{ij}\right) 
		\psi_j,
		\end{aligned}
		\end{align}
		and the problem becomes similar to that of a quantum dot with a time-dependent perturbation~\cite{Kravtsov-DL,Skvortsov-DL,Aleiner-DL1,Aleiner-DL2}.

	We now perform the ensemble averaging for the chaotic ($h\neq V$) and integrable ($h=V$) cases separately.
	For the chaotic ($h \neq V$) case, performing the ensemble average over $h$ and $V$ in Eq.~\ref{Seq:K-1}, we find
	\begin{align}\label{Seq:Ft-0}
	\begin{aligned}
	&\left\langle 
	\exp 
	\left\lbrace 
	\begin{aligned}
	-i\sum_{a=\pm}
	\zeta_a
	\int_0^{t} d t
	\bpsi_i^{a} (t') 	
	\left[  h_{ij}+  \phi^a(t')V_{ij} \right] 
	\psi_j^{a}(t')
	\end{aligned}	
	\right\rbrace 
	\right\rangle 
	\\
	=&
	\int  \D h
	\dfrac{e^{-\frac{N}{2J^2} \Tr h^2 }}{Z_h}
	\exp 
	\left[ 
	-i\sum_{a=\pm}\zeta_a
	\int_0^{t} d t
	\bpsi_i^{a} (t') 	
	h_{ij}
	\psi_j^{a}(t')
	\right] 
	\int  \D V
	\dfrac{e^{-\frac{N}{2J^2} \Tr V^2 }}{Z_h}
	\exp 
	\left[ 
	-i\sum_{a=\pm}\zeta_a
	\int_0^{t} d t
	\bpsi_i^{a} (t') 	
	\phi^a(t')V_{ij}
	\psi_j^{a}(t')
	\right] 
	\\
	=\,&
	\exp 
	\left\lbrace 
	-\suml{a,b} \frac{J^2}{2N} \zeta_a \zeta_b 
	\int_0^{t} d t_1
	\int_0^{t} d t_2
	\left[ 
	1+
	\phi^a(t_1)
	\phi^b(t_2)
	\right] 
	\bpsi_{i}^{a} (t_1) \psi_{j}^{a} (t_1) 
	\bpsi_{j}^{b} (t_2) \psi_{i}^{b} (t_2)  
	\right\rbrace .
	\end{aligned}	
	\end{align}
	By contrast, for  the integrable ($h = V$) case, the $h-$dependent term  after ensemble average becomes 
	\begin{align}
	\begin{aligned}
	&\left\langle 
		\exp 
		\left\lbrace 
		-
		i \suml{a}
		\zeta_a 
		\int_0^{t} d t'
		\bpsi_{i}^{a} (t')
		\left( 1+\phi^a(t')\right) 
		h_{ij} 
		\psi_{j}^{a} (t')
		\right\rbrace
	\right\rangle 
	\\
	=\,&
	\int \D h
	\dfrac{e^{-\frac{N}{2J^2} \Tr V^2 }}{Z}
	\exp \left[ 
	-
	i \suml{a} \zeta_a
	\int_0^{t} d t'
	\bpsi_{i}^{a} (t')
	\left( 1+\phi^a(t')\right) 
	h_{ij} 
	\psi_{j}^{a} (t')
	\right] 
	\\
	=\,&
	\exp 
	\left(-\suml{a,b} \frac{J^2}{2N} \zeta_a \zeta_b 
	\int_0^{t} d t_1
	\int_0^{t} d t_2
	\left[
	\left( 1+\phi^a(t_1)\right) 
	\left( 1+\phi^b(t_2)\right) 
	\right]
	\bpsi_{i}^{a} (t_1) \psi_{j}^{a} (t_1) 
	\bpsi_{j}^{b} (t_2) \psi_{i}^{b} (t_2)  
	\right) .
	\end{aligned}	
	\end{align}
 	Inserting these expressions into Eq.~\ref{Seq:K-1}, we obtain 
		\begin{align}\label{Seq:K2}
		\begin{aligned}
		K(t)
		=		&
		\int D\phi \frac{e^{-S_0[\phi]}}{Z_{\phi}}
		\int \D (\bpsi, \psi)  
		\exp 
		\left\lbrace 
		i\sum_{a} 
		\int_0^{t} d t'
		\bpsi_i^{a} (t') 	
		i \partial_{t'} 
		\psi_i^{a}(t')
		\right.
		\\
		&
		\left.
		-\suml{a,b} \frac{J^2}{2N} \zeta_a \zeta_b 
		\int_0^{t} d t_1
		\int_0^{t}  d t_2
		f^{ab}_{t_1t_2}
		\bpsi_{i}^{a} (t_1) \psi_{j}^{a} (t_1) 
		\bpsi_{j}^{b} (t_2) \psi_{i}^{b} (t_2)  
		\right\rbrace ,
		\end{aligned}	
		\end{align}
	where
		\begin{align}\label{Seq:f}
		f^{ab}_{t_1t_2}
		=
		\begin{dcases}
		1+\phi^a(t_1)\phi^b(t_2),
		&
		h\neq V,
		\\
		\left[ 1+\phi^a(t_1)\right] 
		\left[ 1+\phi^b(t_2)\right] ,
		&
		h=V.
		\end{dcases}
		\end{align}
We would like to emphasize that the seemingly unremarkable difference in this factor between the chaotic and integrable model, eventually gives rise to drastically different SFF behaviors consistent with ergodic and non-ergodic regimes, respectively. 

	We now perform another H.S. transformation and decouple the ensemble-averaging generated quartic interaction. This is a standard procedure in deriving the disordered $\sigma$-model in both non-interacting and interacting cases (see, e.g., Ref.~\cite{Kamenev-book}):
		\begin{align}\label{Seq:K3}
		\begin{aligned}
		K(t)
		=&
		\frac{1}{Z_{\phi}Z_Q}\int D\phi e^{-S_{0}[\phi]}
		\int \D Q
		\exp
		\left( 
		-
		\frac{N}{2J^2}
		\sum_{ab}
		\int_0^{t} d t_1
		\int_0^{t} d t_2
		f^{ab}_{t_1t_2}
		Q^{ab}_{t_1t_2}
		Q^{ba}_{t_2t_1}
		\right) 
		\\
		\times&
		\int \D (\bpsi, \psi)  
		\exp 
		\left\lbrace 
		\begin{aligned}
		i\sum_{a} 
		\int_0^{t} d t_1
		\int_0^{t} d t_2\,
		\bpsi_i^{a}(t_1) 
		\left[ 
		\delta_{ab}\delta_{t_1t_2} i\partial_{t_2}
		+i f^{ab}_{t_1t_2}Q^{ab}_{t_1t_2}\zeta_b
		\right] 
		\psi_i^{b}(t_2)
		\end{aligned}	
		\right\rbrace,
		\\
		Z_Q
		=&
		\int \D Q
		\exp
		\left( 
		-
		\frac{N}{2J^2}\sum_{ab}
		\int_0^{t} d t_1
		\int_0^{t} d t_2
		f^{ab}_{t_1t_2}
		Q^{ab}_{t_1t_2}
		Q^{ba}_{t_2t_1}
		\right)  .
		\end{aligned}	
		\end{align}
		Here $Q$ is a Hermitian matrix satisfying
			\begin{align}
			\begin{aligned}
			Q^{ab}_{t_1+t,t_2}
			=
			-Q^{ab}_{t_1,t_2},
			\qquad
			Q^{ab}_{t_1,t_2+t}
			=
			-Q^{ab}_{t_1,t_2}.
			\end{aligned}
			\end{align}
	
	After integrating out the fermionic field in the equation above, we find that the SFFs of the two ensembles can be expressed as the following $\sigma$-model:
		\begin{align}\label{Seq:K5}
		\begin{aligned}
		&K(t)=\,
		\frac{1}{Z_{\phi}Z_Q}
		\int D\phi e^{-S_{0}[\phi]}
		\int \D Q \exp \left( - S[Q,\phi]\right),
		\\
		&S[Q,\phi]
		=\,	
		\frac{N}{2J^2}
		\sum_{a,b,a',b'}
		\int_{t_1,t_2,t_1',t_2'}
		Q^{ab}_{t_1t_2}
		F^{ba,a'b'}_{t_2t_1,t_1't_2'}
		Q^{b'a'}_{t_2't_1'}
		-
		N
		\Tr 
		\ln
		\left[
		i\zeta_a  \delta_{ab}\delta_{t_1t_2}\partial_{t_2}
		+
		i \sum_{a'b'}\int_{t_1',t_2'}F^{ab,b'a'}_{t_1t_2,t_2' t_1'}Q^{a'b'}_{t_1',t_2'}
		\right]	
		+
		\text{const.}
	    .
		\end{aligned}
		\end{align}
		Here we have employed the notation $\int_{t_1}=\int_0^{t} dt_1$, and $S_0[\phi]$ is defined in Eq.~\ref{Seq:K-1}.
		The kernel $F$ is defined as
		\begin{align}\label{Seq:F}
		 F^{ab,b'a'}_{t_1t_2,t_2't_1'}
		=\delta_{aa'}\delta_{bb'}\delta_{t_1t_1'}\delta_{t_2t_2'} f^{ab}_{t_1t_2},
		\end{align}
		where $f^{ab}_{t_1t_2}$ is given by Eq.~\ref{Seq:f}.

We can also rewrite the theory in the Fourier basis, 
		\begin{align}
		\begin{aligned}
%		\psi^a_n
%		=
%		\int_0^{t}
%		dt' \,
%		\psi^a (t') e^{i \e_n t'},
%		\qquad
%		\\
		\qquad
		\phi^a_m
		=
		\int_0^{t}
		dt' \,
		\phi^a(t') e^{i \ww_m t'},
		\qquad
		Q^{ab}_{nm}
		=
		\int_0^{t} dt_1\int_0^{t} dt_2 \,
		Q^{ab}_{t_1t_2}e^{i \e_n t_1-i \e_m t_2}.
		\end{aligned}
		\end{align}
	Here the fermionic (bosonic) Matsubara frequency is defined as 
	$\e_n=2\pi (n+\frac{1}{2})/t$ ($\ww_m=2\pi m/t$).
	In the Fourier representation, the actions assume the forms
		\begin{align}\label{Seq:K5-FT}
		\begin{aligned}
		&S_{0}[\phi]
		=
		-\sum_{a=\pm} \frac{i\zeta_a }{2U}  \frac{1}{t} \sum_{m}  \phi_{m}^a  \phi_{-m}^a,
		\\
		&S[Q,\phi]
		=\,	
		\frac{N}{2J^2}
		\frac{1}{t^4}
		\sum_{aba'b'}
		\sum_{n,m,n',m'}
		Q^{ab}_{nm}
		F^{ba,a'b'}_{mn,n'm'}
		Q^{b'a'}_{m'n'}
		-
		N
		\Tr 
		\ln
		\left[
		\zeta_a \e_n t\delta_{nn'} \delta_{ab}
		+
		i\frac{1}{t^2} \sum_{a',b'}\sum_{m,m'}F^{ab,b'a'}_{nn',m'm}Q^{a'b'}_{mm'}
		\right]	
		+
		\text{const.},
		\end{aligned}
		\end{align}
		where
		\begin{align}
		F^{ba,a'b'}_{mn,n'm'}
		=\delta_{aa'}\delta_{bb'} f^{ba}_{m-m',n'-n},
		\qquad
		f^{ba}_{mn}
		=
		\begin{dcases}
		t\delta_{n,0}  t \delta_{m,0}+\phi^a_n\phi^b_m,
		&
		h\neq V,
		\\
		\left( t\delta_{n,0}+\phi^a_n\right) 
		\left( t\delta_{m,0}+\phi^b_m\right) ,
		&
		h=V.
		\end{dcases}
		\end{align}	
	
\section{Time reparametrization approach for the integrable $h=V$ model}\label{Ssec:TR}

For the integrable case, there is a way to map the calculation of the SFF to that of the non-interacting model with $U=0$.
Starting from Eq.~\ref{Seq:K5}, one can apply the time reparametrization:
	\begin{align}\label{Seq:TR}
	\begin{aligned}
%	t_1 \rightarrow \tau_1^a = \int_{0}^{t_1} d t' \left[ 1+\phi^a(t') \right] ,
%	\qquad
%	Q^{ab}_{t_1t_2} \rightarrow \bQ^{ab}_{\tau_1^a ,\tau_2^b }
  	Q^{ab}_{t_1t_2} \rightarrow \bQ^{ab}_{\tau^a(t_1) ,\tau^b(t_2) },
  	\qquad
	\tau^a(t_1) \equiv \int_{0}^{t_1} d t' \left[ 1+\phi^a(t') \right],
	\end{aligned}
	\end{align}
and rewrite the action as
	\begin{align}
	\begin{aligned}
	&S[\bQ,\phi]
	=\,	
	\frac{N}{2J^2}
	\int_0^{t+\phi^a_0} d \tau_1^a
	\int_0^{t+\phi^b_0} d \tau_2^b
	\bQ^{ab}_{\tau_1^a,\tau_2^b}
	\bQ^{ba}_{\tau_2^b,\tau_1^a}
	-
	N
	\Tr 
	\ln
	\left[
	i\zeta_a  \delta_{ab} \delta_{\tau_1^a \tau_2^b} \partial_{\tau_2^b}
	+
	i \bQ^{ab}_{\tau_1^a \tau_2^b}
	\right]	
	+
	\text{const.}.
	\end{aligned}
	\end{align}
Switching to the Fourier basis, the action after time reparametrization takes the form
\begin{align}\label{Seq:S-TR-FT}
\begin{aligned}
&S[\bar{Q},\phi]
=\,	
\frac{N}{2J^2}
\sum_{ab}
\frac{1}{(t+\phi^a_0)(t+\phi^b_0)}
\sum_{n,m}
\bar{Q}^{ab}_{nm}
\bar{Q}^{ba}_{mn}
-
N
\Tr 
\ln
\left[
\zeta_a \bar{\e}_n^a (t+ \phi^a_0)\delta_{nn'} \delta_{ab}
+
i \bar{Q}^{ab}_{nn'}
\right]	
+
\text{const.},
\end{aligned}
\end{align}
where
\begin{align}
\begin{aligned}
	\bar{Q}^{ab}_{nm}
	=
	\int_0^{t+\phi^a_0} d\tau_1^a\int_0^{t+\phi^b_0} d\tau_2^b \,
	\bar{Q}^{ab}_{\tau_1^a \tau_2^b}e^{i \bar{\e}_n^a \tau_1^a-i \bar{\e}_m^b \tau_2^b},
	\qquad
	\bar{\e}_n^a=\frac{2\pi}{t+\phi^a_0} (n+1/2).
\end{aligned}
\end{align}
One can see from Eq.~\ref{Seq:S-TR-FT} that,  
for any pair of integers $n,m$ obeying $\bar{\e}_n^+=-\bar{\e}_m^-$, the time-reparametrized action $S[\bar{Q},\phi]$ is invariant under any $U(2)$ rotation applying to the the following rescaled $Q$ subblock:
\begin{align}
	\begin{bmatrix}
	\bar{Q}^{++}_{nn}/(t+\phi^+_0)
	& 
	\bar{Q}^{+-}_{nm}/\sqrt{(t+\phi^+_0)(t+\phi^-_0)} 
	\\
	\bar{Q}^{-+}_{mn}/\sqrt{(t+\phi^+_0)(t+\phi^-_0)} 
	& 
	\bar{Q}^{--}_{mm}/(t+\phi^-_0)
	\end{bmatrix}
	\rightarrow 
	R
	\begin{bmatrix}
	\bar{Q}^{++}_{nn}/(t+\phi^+_0)
	& 
	\bar{Q}^{+-}_{nm}/\sqrt{(t+\phi^+_0)(t+\phi^-_0)} 
	\\
	\bar{Q}^{-+}_{mn}/\sqrt{(t+\phi^+_0)(t+\phi^-_0)} 
	& 
	\bar{Q}^{--}_{mm}/(t+\phi^-_0)
	\end{bmatrix}
	R^{-1}.
\end{align}

Comparing the action after time reparametrization with the non-interacting action $S[Q,\phi=0]$, we find that the SFF of the integrable model be expressed as
	\begin{align}\label{Seq:Kin}
	\begin{aligned}
	&K(t)=\,
	\frac{1}{\int \D \phi_0 e^{-S_0'[\phi_0]  }}
	\int \D \phi_0 e^{-S_0'[\phi_0]  }
	\left\langle 
	\Tr \exp\left[ -i H_0 \left( t+\phi_0^+\right) \right] 
	\Tr \exp \left[ iH_0 \left( t+\phi_0^-\right) \right] 
	\right\rangle,
	\\
&	S_0'[\phi_0]=-\sum_{a} \frac{i\zeta_a }{2U t} (\phi_0^a)^2  ,
	\end{aligned}
	\end{align}
where 	$H_{0}$ is the non-interacting Hamiltonian with $U=0$, and $\phi^a_0\equiv \int_{0}^{t} d t_1 \phi^a(t_1)$.
Note that $\phi^a_0$ is denoted by $\bar{\phi}^a$ in the main text.
Eq.~\ref{Seq:Kin} can also be easily derived from the definition of SFF: 
\begin{align}
\begin{aligned}
	& K(t)=\sum_{nm} \exp\left( -i E_n t +i E_m t \right) 
	=\sum_{nm} 
	\exp\left\lbrace 
	 -i \left[ E_n^{(0)} +\frac{U}{2}( E_n^{(0)})^2 \right]  t 
	 +i \left[ E_m^{(0)} +\frac{U}{2}  (E_m^{(0)})^2 \right] t
	 \right\rbrace 
	\\
	=\,&
	\frac{1}{\int \D \phi_0 e^{-S_0'[\phi_0]  }}
	\int \D \phi_0 e^{-S_0'[\phi_0]  }
	\sum_{nm} \exp\left( -i E_n^{(0)}  (t+\phi^+_0) +i E_m^{(0)} ( t+\phi^-_0) \right) .
\end{aligned}
\end{align}
Here we have used the fact that  $E$ (the eigenenergy of $H$) is related to $E^{(0)}$ (the eigenenergy of $H_0$) by $E=E^{(0)} +\frac{U}{2} (E^{(0)})^2$.
The problem therefore is reduced to that of the non-interacting model $H_0$.

In Ref.~\cite{PRL}, it has been found that the noninteracting SFF is given by the following cumulant expansion:
\begin{align}\label{Seq:CumSum-0}
\begin{aligned}
K_0(t)  
=\,&
\exp
\left\lbrace 
\sum_{n}
\frac{N^n}{n!} 
\int_{-\infty}^{\infty} ...\int_{-\infty}^{\infty} 
\left\lbrace 
\prod_{k=1}^{n}
d \e_k
\left[ \suml{a_k}
\ln \left( 1+e^{-i \zeta_{a_k}  \e_k t }\right) \right] 
\right\rbrace 
R_n^{\msf{con}}(\e_1,...,\e_n)
\right\rbrace .
\end{aligned}
\end{align}
Through a similar derivation, one can show that the SFF of the integrable ensemble investigated in the current paper is given by
	\begin{align}\label{Seq:CumSum}
	\begin{aligned}
	K(t)  
	=\,&
	\frac{1}{\int \D \phi_0 e^{-S_0'[\phi_0]  }}
	\int \D \phi_0 e^{-S_0'[\phi_0]   }
	\\
	&\times
	\exp
	\left\lbrace 
	\sum_{n}
	\frac{N^n}{n!} 
	\int_{-\infty}^{\infty} ...\int_{-\infty}^{\infty} 
	\left\lbrace 
	\prod_{k=1}^{n}
	d \e_k
	\left[ \suml{a_k}
	\ln \left( 1+e^{-i \zeta_{a_k}  \e_k (t+\phi^{a_k}_0)}\right) \right] 
	\right\rbrace 
	R_n^{\msf{con}}(\e_1,...,\e_n)
	\right\rbrace .
	\end{aligned}
	\end{align}
Here $R_n^{\msf{con}}$ represents the connected part of the $n$-point level correlation function $R_n$ for the GUE matrix $h$:
\begin{align}\label{Seq:Rn}
	R_n(\e_1,\e_2,...,\e_n)=\left\langle  \nu(\e_1)\nu(\e_2)...\nu(\e_n)\right\rangle,
\end{align}
where $\nu(\ww)=\frac{1}{N}\sum_{i=1}^N \delta(\ww-\e_i)$ is the density of states.
Inserting the explicit expression for $R_n^{\msf{con}}$~\cite{Mehta} into the equation above, one can solve the SFF of the integrable model in a similar way as the noninteracting model (see Sec.~\ref{sec:nonint} and Ref.~\cite{PRL}).
In particular, each term in the exponent in Eq.~\ref{Seq:CumSum} can be deduced from that of the noninteracting theory by shifting the time variable $t \rightarrow t+\phi^{a_k}_0$.
After summing over all contributions, the final result can be obtained by averaging over fluctuations of the bosonic field $\phi_0$ governed by the action $S_0'[\phi_0]$ (Eq.~\ref{Seq:Kin}).

In Fig.~2 in the main text, we plot the numerical result of the SFF of the integrable ensemble for $N=10$ and $J=1$. %and limited to small ``integrable interaction'' strength $U$, which is the regime of interest to us. 
For weak interaction strength, the non-interacting ramp is only slightly modified and remains exponential. 
This is due to the diffuson-type collective modes that do not acquire a ``dephasing mass,'' in sharp contrast to the chaotic regime discussed below. 
We see that in both the non-interacting and our $h=V$ integrable case, the energy levels are uncorrelated when their energy separation is of the order of many-body level spacing.
Note that for stronger interaction strength, the ramp disappears all together and the SFF shows a non-universal slope down to the plateau.
In this case, the disconnected SFF gives rise to a non-negligible contribution to the slope even around $t \gtrsim N$, which is responsible for the late plateau time.

\section{Saddle point}	

\subsection{Saddle points}

In the following, we assume a weak enough interaction strength $U$ and perform a perturbative calculation for both the chaotic and integrable ensembles. The new result here is that we are able to obtain an interacting saddle point, in contrast to the non-interacting  saddle points usually used in the $\sigma$-model approach.
Our analytical results rely on a straightforward perturbation theory with respect to the H.S. field $\phi$.
We point out that interacting saddle points have been considered before
by Kravtsov et al.~\cite{Kravtsov-DL,Skvortsov-DL} in the context of driven quantum dot,
by Kamenev and Andreev~\cite{KamenevAndreev} to address Altshuler-Aronov corrections in disordered metals,
and numerically by Saad et al~\cite{SSS} for the SFF of the SYK$_4$ model.

%For the integrable case, we also compare the result obtained from perturbation calculation below with that from the aforementioned time reparametrization approach discussed in Sec.~\ref{Ssec:TR}.

Starting from Eq.~\ref{Seq:K5}, we first perform the integration over the matrix field $Q$
\begin{align}
	K_{\phi} \equiv \int \D Q \exp \left( - S[Q,\phi]\right),
\end{align}
by approximating it with the contributions from saddle point configurations $\Qsp$ and the Gaussian fluctuations $\delta Q$ around them, for any real bosonic field $\phi$.

The saddle point equation for the matrix field $Q$ can be obtained by setting the variation of the action $S[Q,\phi]$ with respect to $Q$ to zero:
	\begin{align}	\label{Seq:SPEQ}
%	&\frac{\delta S[Q,\phi]}{\delta Q}\bigg\lvert_{Q=Q_{sp}}=0
%	\Rightarrow
	Q_{sp}
	=
	J^2
	\left( 
	\sigma^3 \partial_{t}
	+
	F Q_{sp}
	\right)^{-1}.
	\end{align}
	Here $\sigma^3$ indicates the third Pauli matrix in the replica space.
	We solve this saddle point equation perturbatively up to order $\phi^2$. In the following, we use $\Qsp^{(n)}$ ($F^{(n)}$) to denote  the $n$-th order term in the expansion of $\Qsp$ ($F$)  in powers of $\phi$:
	\begin{align}
	\begin{aligned}
	Q_{sp}=\Qsp^{(0)} +\Qsp^{(1)}+\Qsp^{(2)}+O(\phi^3),
	\qquad
	F=F^{(0)}+F^{(1)}+F^{(2)}+O(\phi^3).
	\end{aligned}
	\end{align}
	Inserting this equation into Eq.~\ref{Seq:SPEQ}, one obtains at each order in $\phi$ the following saddle point equations:
	\begin{subequations}
	\begin{align}
	&\label{Seq:SP0}
	{Q}_{sp}^{(0)}
	=
	J^2
	\left( 
	\sigma^3 \partial_{t}
	+
	Q_{sp}^{(0)}
	\right) ^{-1} ,
%	\equiv 
%	J^2 G_0,
	\\
	&\label{Seq:SP1}
	Q_{sp}^{(1)}
	=
	-\frac{1}{J^2}
	{Q}_{sp}^{(0)}
	\left( F^{(1)}\Qsp^{(0)}+\Qsp^{(1)}\right) 
	{Q}_{sp}^{(0)},
	\\
	&\label{Seq:SP2}
	Q_{sp}^{(2)}
	=
	-\frac{1}{J^2}
	{Q}_{sp}^{(0)}
	\left( F^{(2)}\Qsp^{(0)}+F^{(1)}\Qsp^{(1)}+\Qsp^{(2)}\right) 
	{Q}_{sp}^{(0)}
	+
	\frac{1}{J^4}
	{Q}_{sp}^{(0)}
	\left( F^{(1)}\Qsp^{(0)}+\Qsp^{(1)}\right) 
	{Q}_{sp}^{(0)}
	\left( F^{(1)}\Qsp^{(0)}+\Qsp^{(1)}\right)
	{Q}_{sp}^{(0)}.
	\end{align}
	\end{subequations}

	At order $\phi^0$, the saddle point Eq.~\ref{Seq:SP0} reduces to that of the non-interacting theory.
	Assuming that the leading order saddle point $Q_{sp}^{(0)}$ is diagonal in the replica and Matsubara frequency spaces, we find
		\begin{align}\label{Seq:la}
		\begin{aligned}
		&
		(Q_{sp}^{(0)})_{nm}^{ab}=\delta_{nm}\delta_{ab}\lambda_{n}^{a}t,
		\\
		&
		\lambda_{n}^{a}
		=
		\frac{1}{2}
		\left( 
		i\zeta_a\e_n
		+
		s_n^a
		\sqrt{4J^2 -(\e_n)^2}
		\right).
%		\approx
%		J s_n^a ,
%		\qquad
%		s_n^a=\pm 1,
		\end{aligned}
		\end{align}
	Here the sign factor $s_n^a$ can take values of $+1$ and $-1$ when $|\e_n|<2J$, and is determined by the fact that $\lambda_n^a$
    should approach zero in the limit $J\ll |\e_n|$ when $|\e_n|>2J$.
    
	Using this result of $\Qsp^{(0)}$ and the explicit expression for $F$ (Eqs.~\ref{Seq:f} and~\ref{Seq:F}), the solution to the $\phi^1$ order saddle point equation (Eq.~\ref{Seq:SP1}) can therefore be easily calculated from
	\begin{align}	\label{Seq:SP1-FT}
	\begin{aligned}
	(Q_{sp}^{(1)})^{aa'}_{nn'}
%	=
%	-\frac{1}{t^a t^{a'}} \frac{1}{J^2}
%	(\Qsp^{(0)})^{aa}_{nn}
%	\left( F^{(1)}\Qsp^{(0)}+\Qsp^{(1)}\right)^{aa'}_{nn'} (\Qsp^{(0)})^{a'a'}_{n'n'}
%	\\
	=&
	-\frac{1}{J^2}
	\lambda_{n}^{a} 
	\left( 
	\frac{1}{t} \suml{m}
	(F^{(1)})^{aa,aa}_{nn',mm}\lambda^{a}_{m}\delta_{aa'}
	+
	(\Qsp^{(1)})^{aa'}_{nn'}\right)
	\lambda_{n'}^{a'} ,
%	\\
%	=&
%	-\frac{1}{J^2}
%	\lambda_{n}^{a} 
%	\left( \phi^{a}_{n-n'} (\lambda^{a}_{n}+\lambda^{a}_{n'})\delta_{aa'}+(\Qsp^{(1)})^{aa'}_{nn'}\right)
%	\lambda_{n'}^{a'} 
	\end{aligned}
	\end{align}
	where 
	\begin{align}
	(F^{(1)})^{aa,aa}_{nn',mm}
	=
	(f^{(1)})^{aa}_{n-m,m-n'}
	=
	\begin{dcases}
	0,
	&
	h\neq V,
	\\
	t \delta_{mn'}\phi^a_{n-m}+t\delta_{mn}\phi^a_{m-n'},
	&
	h=V.
	\end{dcases}
	\end{align}	
	 In an analogous way, one can calculate $\Qsp^{(2)}$ by inserting the results of $\Qsp^{(0)}$ and $\Qsp^{(1)}$ into Eq.\ref{Seq:SP2} which can be further simplified as
	 \begin{align}	\label{Seq:SP2-FT}
	 \begin{aligned}
	 (Q_{sp}^{(2)})^{aa'}_{nn'}
%	 =&
%	 -\frac{1}{J^2}
%	 \frac{1}{t^a t^{a'}}
%	 (\Qsp^{(0)})^{aa}_{nn}
%	 \left( F^{(2)}\Qsp^{(0)}+F^{(1)}\Qsp^{(1)}+\Qsp^{(2)}\right)^{aa'}_{nn'} (\Qsp^{(0)})^{a'a'}_{n'n'}
%	 \\
%	 +&
%	 \frac{1}{J^4}\frac{1}{t^a (t^{a''})^2 t^{a'}}
%	 (\Qsp^{(0)})^{aa}_{nn}\left( F^{(1)}\Qsp^{(0)}+\Qsp^{(1)}\right)^{aa''}_{nn''} (\Qsp^{(0)})^{a''a''}_{n''n''}
%	 \left( F^{(1)}\Qsp^{(0)}+\Qsp^{(1)}\right)^{a''a'}_{n''n'}(\Qsp^{(0)})^{a'a'}_{n'n'}
%	 \\
	 =&
	 -\frac{1}{J^2}
	 \la^{a}_{n}
	 \left(
	 \frac{1}{t}\suml{m}
	 (F^{(2)})^{aa,aa}_{nn',mm}\lambda^a_m \delta_{aa'}
	 +
	 \frac{1}{t^2}\suml{m,m'}
	 (F^{(1)})^{aa',a'a}_{nn',m'm}(\Qsp^{(1)})^{aa'}_{mm'}
	 +(\Qsp^{(2)})^{aa'}_{nn'}
	 \right) 
	 \la^{a'}_{n'}
	 \\
	 +&
	 \frac{1}{J^4}\frac{1}{t}
	 \suml{m,a''}
	 \la^{a}_{n}\left( F^{(1)}\Qsp^{(0)}+\Qsp^{(1)}\right)^{aa''}_{nm} \la^{a''}_{m}
	 \left( F^{(1)}\Qsp^{(0)}+\Qsp^{(1)}\right)^{a''a'}_{mn'}\la^{a'}_{n'}
	 \\
	 =&
	 -\frac{1}{J^2}
	 \la^{a}_{n}
	 \left(
	 \frac{1}{t}\suml{m}
	 (F^{(2)})^{aa,aa}_{nn',mm}\lambda^a_m
	  \delta_{aa'}
	 +
	 \frac{1}{t^2}\suml{m,m'}
	 (F^{(1)})^{aa',a'a}_{nn',m'm}(\Qsp^{(1)})^{aa'}_{mm'}
	 +(\Qsp^{(2)})^{aa'}_{nn'}
	 \right) 
	 \la^{a'}_{n'}
	 \\
	 +&
	 \frac{1}{J^4}\frac{1}{t}
	 \suml{m,a''}
	 \la^{a}_{n}
	 \frac{(Q_{sp}^{(1)})^{aa''}_{nm}}{\la^{a}_{n}\la^{a''}_{m}/J^2}
	 \la^{a''}_{m}
	 \frac{(Q_{sp}^{(1)})^{a''a'}_{mn'}}{\la^{a''}_{m}\la^{a'}_{n'}/J^2}
	 \la^{a'}_{n'},
	 \end{aligned}
	 \end{align}
	 where
	 \begin{align}
	 \begin{aligned}
	 &(F^{(2)})^{aa,aa}_{nn',mm}
	 =
	 (f^{(2)})^{aa}_{n-m,m-n'}
	 =
	 \phi^a_{n-m}\phi^a_{m-n'},
	 \\
	 &(F^{(1)})^{aa',a'a}_{nn',m'm}
	 =
	 (f^{(1)})^{aa'}_{n-m,m'-n'}
	 =
	 \begin{dcases}
	 0,
	 &
	 h\neq V,
	 \\
	 t\delta_{nm} \phi^{a'}_{m'-n'}
	 +
	 t\delta_{n'm'} \phi^{a}_{n-m},
	 &
	 h=V.
	 \end{dcases}
	 \end{aligned}
	 \end{align}
	In the second equality of Eq.~\ref{Seq:SP2-FT}, we have made use of the fact that $\Qsp^{(1)}$ is the solution to saddle point Eq.~\ref{Seq:SP1} and satisfies
	 \begin{align}	
	 \begin{aligned}
%	 &(Q_{sp}^{(1)})^{aa'}_{nn'}
%	 =
%	 -\frac{1}{t^a t^{a'}} \frac{1}{J^2}
%	 (\Qsp^{(0)})^{aa}_{nn}
%	 \left( F^{(1)}\Qsp^{(0)}+\Qsp^{(1)}\right)^{aa'}_{nn'} (\Qsp^{(0)})^{a'a'}_{nn'}
%	 =
%	 -\frac{1}{J^2}
%	 \lambda_{n}^{a} 
%	 \left( \phi^{a}_{n-n'} (\lambda^{a}_{n}+\lambda^{a}_{n'})\delta_{aa'}+(\Qsp^{(1)})^{aa'}_{nn'}\right)
%	 \lambda_{n'}^{a'} 
%	 \\
%	 &\Rightarrow 
	 \left( F^{(1)}\Qsp^{(0)}+\Qsp^{(1)}\right)^{aa'}_{nn'}
	 =
	 -\frac{J^2}{\la^{a}_{n}\la^{a'}_{n'}}(Q_{sp}^{(1)})^{aa'}_{nn'}.
%	 =
%	 \delta_{a,a'}
%	 \dfrac{1}{1+\lambda^{a}_{n}\lambda^{a}_{n'}/J^2}\phi^{a}_{n-n'} (\lambda^{a}_{n}+\lambda^{a}_{n'})
	 \end{aligned}
	 \end{align}

	 Solving Eq.~\ref{Seq:SP1-FT} and subsequently Eq.\ref{Seq:SP2-FT}, we find that,
	for the chaotic $h\neq V$ case, the saddle point solution takes the following  form in terms of $\lambda_n^a$ (Eq.~\ref{Seq:la}):
	\begin{align}\label{Seq:QSPa}
	\begin{aligned}
	&(Q_{sp}^{(0)})_{nn'}^{aa'}=\delta_{aa'}\delta_{nn'}\lambda_{n}^{a}t,
	\\
	&\Qsp^{(1)}=0,
	\\
	& (Q_{sp}^{(2)})^{aa'}_{nn'}
	=
	\delta_{aa'} 
	\frac{-\la^{a}_{n} \la^{a}_{n'}/J^2}{1+\la^{a}_{n}\la^{a}_{n'}/J^2}
	\frac{1}{t} \sum_{m} \phi^a_{n-m} \phi^a_{m-n'}
	\lambda^a_m.
	\end{aligned}
	\end{align}
	On the other hand, for the integrable $h=V$ case, up to order $\phi^2$, the saddle point solution is given by
	\begin{align}\label{Seq:QSPb}
	\begin{aligned}
	& (Q_{sp}^{(0)})_{nn'}^{aa'}=\delta_{aa'}\delta_{nn'}\lambda_{n}^{a}t,
	\\
	& (Q_{sp}^{(1)})^{aa'}_{nn'}
	=
	\delta_{aa'}
	\dfrac{-\lambda^{a}_{n}\lambda^{a}_{n'}/J^2}{1+\lambda^{a}_{n}\lambda^{a}_{n'}/J^2}\phi^{a}_{n-n'} (\lambda^{a}_{n}+\lambda^{a}_{n'}),
	\\
	&(Q_{sp}^{(2)})^{aa'}_{nn'}
	=
	\delta_{aa'} \frac{-\la^{a}_{n}	\la^{a}_{n'}/J^2 }
	{1+\la^{a}_{n}	\la^{a}_{n'}/J^2}
	\frac{1}{t} \sum_{m} \phi^a_{n-m} \phi^a_{m-n'} 
	\left\lbrace 
	\begin{aligned}
	&
	\lambda^a_m	
	+
	\left[ 
	\dfrac{-\lambda^{a}_{m}\lambda^{a}_{n'}/J^2}{1+\lambda^{a}_{m}\lambda^{a}_{n'}/J^2} (\lambda^{a}_{m}+\lambda^{a}_{n'})
	+
	\dfrac{-\lambda^{a}_{m}\lambda^{a}_{n}/J^2}{1+\lambda^{a}_{m}\lambda^{a}_{n}/J^2}
	(\lambda^{a}_{m}+\lambda^{a}_{n})
	\right] 
	\\
	&
	-
	\frac{1}{J^2}\la^{a}_{m}
	\dfrac{\lambda^{a}_{n}+\lambda^{a}_{m}}{1+\lambda^{a}_{n}\lambda^{a}_{m}/J^2} 
	\dfrac{\lambda^{a}_{m}+\lambda^{a}_{n'}}{1+\lambda^{a}_{m}\lambda^{a}_{n'}/J^2} 
	\end{aligned}
	\right\rbrace.
	\end{aligned}	
	\end{align}
Note that here the saddle point solutions are written in terms of $\lambda_n^a$ defined in Eq.~\ref{Seq:la} and depend on the sign factor $s_n^a$, which can take both the values of $+1$ and $-1$ when $|\e_n|<2J$.
As a result, we find a series of different $\phi$-dependent saddle points corresponding to different choices of $\left\lbrace s_n^a \right\rbrace $.

\subsection{Saddle point action}

Now, we will evaluate the action at various saddle points, by inserting Eqs.~\ref{Seq:QSPa} and.~\ref{Seq:QSPb} into the expression for the action $S[Q,\phi]$ in Eq.~\ref{Seq:K5}. To evaluate the $\Tr\ln$ term, we apply a second order Taylor expansion:
\begin{align}\label{Seq:TE}
\begin{aligned}
	S[\Qsp,\phi]
	=&
	S[\Qsp,0]
	+
	\suml{a}\int_{0}^{t} dt_1
	\frac{\delta S[\Qsp,\phi]}{\delta \phi^a_{t_1}}\bigg|_{\phi=0} \phi^a_{t_1}
	+
	\frac{1}{2}
	\suml{a,b}\int_{0}^{t} dt_1 \int_{0}^{t} dt_2
	\frac{\delta^2 S[\Qsp,\phi]}{\delta \phi^a_{t_1}\delta \phi^b_{t_2}}\bigg|_{\phi=0} \phi^a_{t_1}\phi^b_{t_2}
	+O(\phi^3).
\end{aligned}
\end{align}
Here $S[\Qsp,0]$ is the saddle point action for the non-interacting theory, and has been evaluated in Ref.~\cite{PRL}:
	\begin{align}\label{Seq:S0-1}
	\begin{aligned}
	S[\Qsp,0]
	=\,	
	N
	\sum_{a,n}
	\left[
	\frac{1}{2J^2}
	\left( \lambda_{n}^{a}\right)^2
	-
	\ln
	\left( 
	\zeta_a \e_n 
	+
	i\lambda_{n}^{a} 
	\right)
	\right]
	=
	\frac{N}{J}	
	\sum_{a,n}
	i\zeta_a 
	s_n^a
	\e_n
	+
	O(\e_n^2)
	+
	\text{const.}.
\end{aligned}
\end{align}
In the second equality, we have applied an expansion in terms of energy, and kept only the leading order terms.

The functional derivative $\delta S[\Qsp,\phi]/\delta \phi^a_{t_1}$ can be straightforwardly computed using the fact that matrix field $Q$ is at its saddle point:
\begin{align}\label{Seq:dS1}
\begin{aligned}
&\frac{\delta S[\Qsp,\phi]}{\delta\phi^{a}_{t_1}}
=\,	
N
\suml{a'b'}
\int_0^{t} d t_1'
\int_0^{t} d t_2'
\left\lbrace 
\frac{1}{2J^2}
(\Qsp)^{a'b'}_{t_1',t_2'}
(\Qsp)^{b'a'}_{t_2',t_1'}
-
\left[ 
\left( 
\sigma^3 \partial_{t}
+
F \Qsp
\right) 
^{-1}
\right]^{a'b'}_{t_1',t_2'}
(\Qsp)^{b'a'}_{t_2',t_1'}
\right\rbrace 
\frac{\partial f^{a'b'}_{t_1't_2'}}{ \partial \phi^{a}_{t_1}}
\\
=\,&
-\frac{N}{2J^2}
\suml{a'b'}
\int_0^{t} d t_1'
\int_0^{t} d t_2'
(\Qsp)^{a'b'}_{t_1't_2'}
(\Qsp)^{b'a'}_{t_2't_1'}
\frac{\partial f^{a'b'}_{t_1't_2'}}{ \partial \phi^{a}_{t_1}},
\end{aligned}
\end{align}
where we have utilized the saddle point Eq.~\ref{Seq:SPEQ}.
Substituting the explicit expression for $f$ (Eq.~\ref{Seq:f}) into Eq.~\ref{Seq:dS1}, we find
\begin{align}
\begin{aligned}
\frac{\delta S[\Qsp,\phi]}{\delta\phi^{a}_{t_1}}\bigg|_{\phi=0}
=&
\begin{cases}
-\frac{N}{J^2}
\suml{b}
\int_0^{t} d t_2
(\Qsp)^{ab}_{t_1t_2}
(\Qsp)^{ba}_{t_2t_1}
\phi^b_{t_2}\bigg|_{\phi=0},
& h\neq V,
\\
-\frac{N}{J^2}
\suml{b}
\int_0^{t} d t_2
(\Qsp)^{ab}_{t_1t_2}
(\Qsp)^{ba}_{t_2t_1}
\left( 1+\phi^b_{t_2} \right)\bigg|_{\phi=0},
& h=V,
\end{cases}
\\
=&
\begin{cases}
0,
&
 h\neq V,
\\
-\frac{N}{J^2}
\suml{b}
\int_0^{t} d t_2
(\Qsp^{(0)})^{ab}_{t_1t_2}
(\Qsp^{(0)})^{ba}_{t_2t_1},
& 
h=V.
\end{cases}
\end{aligned}
\end{align}

Starting from Eq.~\ref{Seq:dS1}, and using the fact that $f^{ab}_{t_1t_2}=f^{ba}_{t_2t_1}$, one can obtain the second derivative of the action with respect to $\phi$:
\begin{align}
\begin{aligned}
&\frac{\delta^2S[\Qsp,\phi]}{\delta\phi^{a}_{t_1}\delta\phi^{b}_{t_2}}
=
-\frac{N}{2J^2}
\suml{a'b'}
\int_0^{t} d t_1'
\int_0^{t} d t_2'
(\Qsp)^{a'b'}_{t_1't_2'}
(\Qsp)^{b'a'}_{t_2't_1'}
\frac{\partial^2 f^{a'b'}_{t_1't_2'}}
{ \partial \phi^{a}_{t_1}\partial \phi^{b}_{t_2}}
-
\frac{N}{J^2}
\suml{a'b'}
\int_0^{t} d t_1'
\int_0^{t} d t_2'
\frac{\partial (\Qsp)^{a'b'}_{t_1't_2'}}{\partial \phi^{b}_{t_2}}
(\Qsp)^{b'a'}_{t_2',t_1'}
\frac{\partial f^{a'b'}_{t_1't_2'}}
{ \partial \phi^{a}_{t_1}}.
\end{aligned}
\end{align}
Evaluating the equation above at $\phi=0$, one arrives at
\begin{align}
\begin{aligned}
&\frac{\delta^2S[\Qsp,\phi]}{\delta \phi^{a}_{t_1} \delta \phi^{b}_{t_2}}\bigg|_{\phi=0}
\\
=&
\begin{cases}
-\frac{N}{J^2}
(\Qsp)^{ab}_{t_1t_2}
(\Qsp)^{ba}_{t_2t_1}
\bigg|_{\phi=0}
-
\frac{N}{J^2}
\suml{a'}
\int_0^{t} d t_1'
\left[ 
\frac{\partial (\Qsp)^{aa'}_{t_1t_1'}}{\partial \phi^{b}_{t_2}}
(\Qsp)^{a'a}_{t_1't_1}
+
(\Qsp)^{aa'}_{t_1t_1'}
\frac{\partial (\Qsp)^{a'a}_{t_1't_1}}{\partial \phi^{b}_{t_2}}
\right] 
\phi^{a'}_{t_1'}
\bigg|_{\phi=0},
& 
h\neq V,
\\
-\frac{N}{J^2}
(\Qsp)^{ab}_{t_1,t_2}
(\Qsp)^{ba}_{t_2,t_1}
\bigg|_{\phi=0}
-
\frac{N}{J^2}
\suml{a'}
\int_0^{t} d t_1'
\left[ 
\frac{\partial (\Qsp)^{aa'}_{t_1t_1'}}{\partial \phi^{b}_{t_2}}
(\Qsp)^{a'a}_{t_1't_1}
+
(\Qsp)^{aa'}_{t_1t_1'}
\frac{\partial (\Qsp)^{a'a}_{t_1't_1}}{\partial \phi^{b}_{t_2}}
\right] 
(1+\phi^{a'}_{t_1'})
\bigg|_{\phi=0},
& 
h=V,
\end{cases}
\\
=&
\begin{cases}
-\frac{N}{J^2}
(\Qsp^{(0)})^{ab}_{t_1t_2}
(\Qsp^{(0)})^{ba}_{t_2t_1},
& 
h\neq V,
\\
-\frac{N}{J^2}
(\Qsp^{(0)})^{ab}_{t_1,t_2}
(\Qsp^{(0)})^{ba}_{t_2,t_1}
-
\frac{N}{J^2}
\suml{a'}
\int_0^{t} d t_1'
\left[ 
\frac{\partial (\Qsp^{(1)})^{aa'}_{t_1t_1'}}{\partial \phi^{b}_{t_2}}
(\Qsp^{(0)})^{a'a}_{t_1't_1}
+
(\Qsp^{(0)})^{aa'}_{t_1t_1'}
\frac{\partial (\Qsp^{(1)})^{a'a}_{t_1't_1}}{\partial \phi^{b}_{t_2}}
\right] 
\bigg|_{\phi=0},
& 
h=V.
\end{cases}
\end{aligned}
\end{align}

Inserting these results for the functional derivatives of the action with respect to the H.S. field $\phi$ into the Taylor expansion Eq.~\ref{Seq:TE}, we obtain the saddle point action up to order $\phi^2$.
In the chaotic $h\neq V$ case, we have
\begin{align}
\begin{aligned}
S[\Qsp,\phi]
=&
S[\Qsp,0]
-
\frac{N}{2J^2}
\suml{a,b}\int_{t_1,t_2}
(\Qsp^{(0)})^{ab}_{t_1t_2}
(\Qsp^{(0)})^{ba}_{t_2t_1}
\phi^a_{t_1}\phi^b_{t_2}
\\
=&
S[\Qsp,0]
-
\frac{N}{2J^2}
\suml{a,b}
\frac{1}{t^4}
\suml{n_1,n_2,n_1',n_2'}
(\Qsp^{(0)})^{ab}_{n_1,n_2}
(\Qsp^{(0)})^{ba}_{n_2',n_1'}
\phi^a_{n_1'-n_1}\phi^b_{n_2-n_2'}
\\
=&
S[\Qsp,0]
-
\frac{N}{2J^2}
\frac{1}{t^2}
\suml{a}
\suml{m,n}
\left( 
\lambda_n^a \lambda_{n+m}^a
\right) 
\phi^a_{m}\phi^{a}_{-m}.
\end{aligned}
\end{align}
On the other hand, for the integrable $h=V$ case,
\begin{align}
\begin{aligned}
S[\Qsp,\phi]
=&
S[\Qsp,0]
-
\frac{N}{J^2}
\suml{a,b}
\int_{t_1,t_2}
(\Qsp^{(0)})^{ab}_{t_1t_2}
(\Qsp^{(0)})^{ba}_{t_2t_1}
\phi^{b}_{t_2}
-
\frac{N}{2J^2}
\suml{a,b}\int_{t_1,t_2}
(\Qsp^{(0)})^{ab}_{t_1t_2}
(\Qsp^{(0)})^{ba}_{t_2t_1}
\phi^a_{t_1}\phi^b_{t_2}
\\
&
-
\frac{N}{2J^2}
\suml{a,b,c}
\int_{t_1,t_2,t_3}
\left[ 
\frac{\partial (\Qsp^{(1)})^{ac}_{t_1,t_3}}{\partial \phi^{b}_{t_2}}
(\Qsp^{(0)})^{ca}_{t_3,t_1}
+
(\Qsp^{(0)})^{ac}_{t_1,t_3}
\frac{\partial (\Qsp^{(1)})^{ca}_{t_3,t_1}}{\partial \phi^{b}_{t_2}}
\right] 
\phi^{a}_{t_1}\phi^{b}_{t_2}
\\
=&
S[\Qsp,0]
-
\frac{N}{2J^2}
\suml{a,b}
\int_{t_1,t_2}
(\Qsp^{(0)})^{ab}_{t_1t_2}
(\Qsp^{(0)})^{ba}_{t_2t_1}
\phi^a_{t_1}\phi^b_{t_2}
\\
&-
\frac{N}{2J^2}
\suml{a,b}
\int_{t_1,t_2}
\left[ 
2
(\Qsp^{(0)})^{ab}_{t_1t_2}
(\Qsp^{(0)})^{ba}_{t_2t_1}
+
(\Qsp^{(1)})^{ab}_{t_1t_2}
(\Qsp^{(0)})^{ba}_{t_2t_1}
+
(\Qsp^{(0)})^{ab}_{t_1t_2}
(\Qsp^{(1)})^{ba}_{t_2t_1}
\right] 
\phi^{a}_{t_1}
\\
=&
S[\Qsp,0]
-
\frac{N}{2J^2}
\suml{a,b}
\frac{1}{t^4}
\suml{n_1,n_2,n_1',n_2'}
(\Qsp^{(0)})^{ab}_{n_1n_2}
(\Qsp^{(0)})^{ba}_{n_2'n_1'}
\phi^a_{n_1'-n_1}\phi^b_{n_2-n_2'}
\\
&
-
\frac{N}{2J^2}
\suml{a,b}
\frac{1}{t^3}
\suml{n_1,n_2,n_2'}
\left[ 
2
(\Qsp^{(0)})^{ab}_{n_1,n_2}
(\Qsp^{(0)})^{ba}_{n_2,n_1'}
+
(\Qsp^{(1)})^{ab}_{n_1,n_2}
(\Qsp^{(0)})^{ba}_{n_2,n_1'}
+
(\Qsp^{(0)})^{ab}_{n_1,n_2}
(\Qsp^{(1)})^{ba}_{n_2,n_1'}
\right] 
\phi^{a}_{n_1'-n_1}
% \\
% =&
% S[\Qsp,0]
% -
% \frac{N}{2J^2}\frac{1}{t^2}
% \suml{a}
% \suml{n,n'}
% \lambda_n^a \lambda_{n'}^a
% \phi^a_{n'-n}\phi^a_{n-n'}
% \\
% &
% -
% \frac{N}{J^2}\frac{1}{t}
% \suml{a}
% \suml{n}
% (\lambda_{n}^a)^2
% \phi^{a}_{0}
% -
% \frac{N}{2J^2}
% \frac{1}{t^2}
% \suml{a}
% \suml{n,n'}
% (\Qsp^{(1)})^{aa}_{n,n'}
% \left( 
% \lambda^a_{n'}
% +
% \lambda^a_{n}
% \right) 
% \phi^{a}_{n'-n}
\\
=&
S[\Qsp,0]
-
\frac{N}{J^2}
\frac{1}{t}
\suml{a,n}
(\lambda_{n}^a)^2
\phi^{a}_{0}
-
\frac{N}{2J^2}
\frac{1}{t^2}
\suml{a,n,m}
\left[ 
\lambda_n^a \lambda_{m}^a
+
\dfrac{-\lambda^{a}_{n}\lambda^{a}_{m}/J^2}{1+\lambda^{a}_{n}\lambda^{a}_{m}/J^2}
(\lambda^{a}_{n}+\lambda^{a}_{m})^2
\right] 
\phi^a_{m-n}\phi^a_{n-m}.
\end{aligned}
\end{align}
    We have derived the expression for the saddle point action in terms of $\left\lbrace \lambda_n^a\right\rbrace$ which varies for different choices of $\left\lbrace s_n^a\right\rbrace$. One can see from the results above that the dominate saddle point which minimizes the action $\re S[\Qsp,\phi]$ depends on the configuration of the bosonic field $\phi$ (as well as the infinitesimal increment of $t^a$~\cite{PRL}).

\section{Quadratic fluctuation and diffuson}\label{Ssec:QF}

Let us now consider the contribution from quadratic fluctuations around the saddle points. 
To proceed, we first make the transformation
\begin{align}
	Q \rightarrow \tilde{Q}=FQ.
\end{align}
As will be elucidated in Sec.~\ref{Ssec:diagram}, $\tilde{Q}_{sp} \equiv F\Qsp$ is related to the self-energy of the ensemble averaged fermionic Green's function, while the fluctuation propagator 
$\left\langle \delta \tilde{Q}  \delta \tilde{Q} \right\rangle \equiv \left\langle F\delta Q F \delta Q \right\rangle $ corresponds to the diffuson defined by series of `impurity' ladder diagrams.

After the transformation, we expand in terms of the fluctuations $\delta \tilde{Q}$ around the saddle point $\tilde{Q}_{sp}$ up to the quadratic order:
\begin{align}\label{Seq:SQ}
\begin{aligned}
	S[\tilde{Q}_{sp}+\delta \tilde{Q},\phi]
	=\,	&
	\frac{N}{2J^2}
	\Tr (\tilde{Q}_{sp}+\delta \tQ) F^{-1} (\tilde{Q}_{sp}+\delta \tQ)
	-
	N
	\Tr 
	\ln
	\left( 
	i \sigma^3 \partial_{t}
	+
	i(\tQ_{sp}+\delta \tQ)
	\right) 
	+
	\text{const.}
	\\
	=&
	S[\tilde{Q}_{sp},\phi]
	+
	\frac{N}{2J^2}
	\Tr \delta \tQ F^{-1} \delta \tQ
	+
	\frac{N}{2}
	\Tr 
	\left( 
	G\delta \tQ G\delta \tQ
	\right) 
	\\
	=\,&
	S[\tilde{Q}_{sp},\phi]
	+
	\frac{N}{2}
	\sum_{a,b,a',b'} \int_{t_1,t_2,t_1',t_2'}
	\delta \tQ^{a'b'}_{t_1't_2'} 
	\left[ 
	\frac{1}{J^2}
	(F^{-1})^{b'a',ab}_{t_2't_1',t_1t_2}
	+
	G^{aa'}_{t_1t_1'} G^{b' b}_{t_2't_2}
	\right] 
	\delta \tQ^{ba}_{t_2t_1}.
\end{aligned}
\end{align}
$G$ is defined as
\begin{align}\label{Seq:G}
	G
	\equiv
	\left(  
	\sigma^3\partial_t
	+
	\tilde{Q}_{sp}
	\right)^{-1}
	=
	\left(  
	\sigma^3\partial_t
	+
	F {Q}_{sp}
	\right)^{-1}
	=
	\frac{1}{J^2}\Qsp,
\end{align}
where in the last equality we have employed the saddle point Eq.~\ref{Seq:SPEQ}.

For simplicity, we define the kernel $M$ as
	\begin{align}\label{Seq:M}
	\begin{aligned}
	&M^{b'a',ab}_{t_2't_1',t_1t_2}
	=
	\frac{1}{J^2}
	(F^{-1})^{b'a',ab}_{t_2't_1',t_1t_2}
	+
	G^{aa'}_{t_1t_1'} G^{b' b}_{t_2't_2}
	=
	\frac{1}{J^2}
	(F^{-1})^{b'a',ab}_{t_2't_1',t_1t_2}
	+
	\frac{1}{J^4}
	({Q}_{sp})^{aa'}_{t_1t_1'} ({Q}_{sp})^{b' b}_{t_2't_2},
	\end{aligned}
	\end{align} 
	and rewrite the action for quadratic fluctuations around $\tilde{Q}_{sp}$ as
	\begin{align}\label{Seq:SQ3}
	\begin{aligned}
	\delta S[\delta \tilde{Q},\phi]
	=\,&
	\frac{N}{2}
	\sum_{a,b,a',b'} \int_{t_1,t_2,t_1',t_2'}
	\delta \tQ^{a'b'}_{t_1't_2'} 
	M^{b'a',ab}_{t_2't_1',t_1t_2}
	\delta \tQ^{ba}_{t_2t_1}.
%	\\
%	=\,&
%	S_{sp}
%	+
%	\frac{N}{2}
%	\sum \int
%	\delta Q^{a',b'}_{t_1',t_2'} 
%	F^{b',a';a',b'}_{t_2',t_1';t_1',t_2'}
%	M^{b',a';a,b}_{t_2',t_1';t_1,t_2}
%	F^{b,a;a,b}_{t_2,t_1;t_1,t_2}
%	\delta Q^{b,a}_{t_2,t_1}
	\end{aligned}
	\end{align}
The propagator for the Gaussian fluctuations $\delta \tilde{Q}$ is given by the inverse of the kernel
\begin{align}\label{Seq:DEQ}
\begin{aligned}
	\D^{ab,b'a'}_{t_1t_2,t_2't_1'} 
	\equiv
	N\left\langle (\delta  \tilde{Q} )^{ab}_{t_1t_2}  (\delta  \tilde{Q})^{b'a'}_{t_2't_1'}  \right\rangle 
	=
	(M^{-1})^{ab,b'a'}_{t_1t_2,t_2't_1'}.
\end{aligned}
\end{align}
The contribution from the quadratic fluctuations around ${Q}_{sp}$ to the SFF is proportional to
\begin{align}\label{Seq:IntdQ}
\begin{aligned}
	\int \D \delta \tilde{Q}
	\exp \left( 
	-\frac{N}{2}
	\sum_{a,b,a',b'} \int_{t_1,t_2,t_1',t_2'}
	\delta \tQ^{a'b'}_{t_1't_2'} 
	M^{b'a',ab}_{t_2't_1',t_1t_2}
	\delta \tQ^{ba}_{t_2t_1}
	\right) 
	\propto
	\exp \left( -\Tr \ln M  \right)
	=
	\exp \left( \Tr \ln \D \right).
\end{aligned}
\end{align}
We emphasize that the kernel $M$ depends on the explicit form of $\Qsp$ and takes different forms for different saddle points.

Switching to the Fourier basis, we have
\begin{align}\label{Seq:S2-E}
\begin{aligned}
	\delta S[\delta \tilde{Q},\phi]
	=\,&
	\frac{N}{2}
	\sum_{a,b,a',b'} 
	\frac{1}{t^4}
	\sum_{n_1,n_2,n_1',n_2'}
	\delta \tQ^{a'b'}_{n_1'n_2'} 
	M^{b'a',ab}_{n_2'n_1',n_1n_2}
	\delta \tQ^{ba}_{n_2n_1},
	\\
	M^{b'a',ab}_{n_2'n_1',n_1n_2}
	=\,&
	\frac{1}{J^2}
	(F^{-1})^{b'a',ab}_{n_2'n_1',n_1n_2}
	+
	\frac{1}{J^4}
	({Q}_{sp})^{aa'}_{n_1n_1'} ({Q}_{sp})^{b' b}_{n_2'n_2},
	\\
	\D^{ab,b'a'}_{n_1n_2,n_2'n_1'} 
	\equiv&
	N\left\langle (\delta  \tilde{Q} )^{ab}_{n_1n_2}  (\delta  \tilde{Q})^{b'a'}_{n_2'n_1'}  \right\rangle 
	=
	t^4 (M^{-1})^{ab,b'a'}_{n_1n_2,n_2'n_1'}.
\end{aligned}
\end{align}
Note here $M^{-1}$ in the time and energy representations are defined separately as the inverse of $M$ in the corresponding basis:
\begin{align}
\begin{aligned}
	&\sum_{a''b''}\int_{t_1'',t_2''}
	(M^{-1})^{ab,b''a''}_{t_1t_2,t_2''t_1''}	
	M^{a''b'',b'a'}_{t_1''t_2'',t_2't_1'}
	=
	\delta_{aa'}\delta_{bb'}\delta_{t_1t_1'}\delta_{t_2t_2'},
	\\
	&\sum_{a''b''}\sum_{n_1'',n_2''}
	(M^{-1})^{ab,b''a''}_{n_1n_2,n_2''n_1''}	
	M^{a''b'',b'a'}_{n_1''n_2'',n_2'n_1'}
	=
	\delta_{aa'}\delta_{bb'}\delta_{n_1n_1'}\delta_{n_2n_2'}.
\end{aligned}
\end{align}

In an expansion up to the second order in $\phi$,  the kernel $M$ takes the following forms in the time and energy representation, respectively, 
	\begin{align}
	\begin{aligned}
	M^{b'a',ab}_{t_2't_1',t_1t_2}
	=\,&
	\frac{1}{J^2}
	\left[ 1-(f^{(1)})^{ab}_{t_1t_2}-(f^{(2)})^{ab}_{t_1t_2}+\left(  (f^{(1)})^{ab}_{t_1t_2}\right)^2 
	\right] 
	\delta_{t_1t_1'}\delta_{t_2t_2'}\delta_{aa'}\delta_{bb'}
	\\
	&+
	\frac{1}{J^4} 
	\left[ 
	(\Qsp^{(0)}+\Qsp^{(1)})^{aa'}_{t_1t_1'} (\Qsp^{(0)}+\Qsp^{(1)})^{b' b}_{t_2't_2}
	+
	(\Qsp^{(0)})^{aa'}_{t_1t_1'} (\Qsp^{(2)})^{b' b}_{t_2't_2}
	+
	(\Qsp^{(2)})^{aa'}_{t_1t_1'} (\Qsp^{(0)})^{b' b}_{t_2't_2}
	\right] 
	+
	O(\phi^3),
	\\
	M^{b'a',ab}_{n_2'n_1',n_1n_2}
	=\,&
	\frac{1}{J^2}
	\left[t^2 \delta_{n_1n_1'} \delta_{n_2'n_2} -(f^{(1)})^{ab}_{n_1-n_1',n_2'-n_2}-(f^{(2)})^{ab}_{n_1-n_1',n_2'-n_2}+\left(  (f^{(1)})^2\right)^{ab}_{n_1-n_1',n_2'-n_2}
	\right] \delta_{aa'}\delta_{b'b}
	\\
	&+
	\frac{1}{J^4} 
	\left[ 
	(\Qsp^{(0)}+\Qsp^{(1)})^{aa'}_{n_1n_1'} (\Qsp^{(0)}+\Qsp^{(1)})^{b'b}_{n_2'n_2}
	+
	(\Qsp^{(0)})^{aa'}_{n_1n_1'} (\Qsp^{(2)})^{b' b}_{n_2'n_2}
	+
	(\Qsp^{(2)})^{aa'}_{n_1n_1'} (\Qsp^{(0)})^{b' b}_{n_2'n_2}
	\right] 
	+
	O(\phi^3).
\end{aligned}
\end{align}
Inserting the previously obtained perturbative expression for $\Qsp$ (Eqs.~\ref{Seq:QSPa} and~\ref{Seq:QSPb}), we find the expression for $M$ in terms of $\lambda_n^a$ (Eq.~\ref{Seq:la}) up to quadratic order in $\phi$.
For the chaotic  case $h\neq V$ case, one has
\begin{align}\label{Seq:M-a}
\begin{aligned}
J^2M^{b',a';a,b}_{n_2',n_1';n_1,n_2}
=\,&
t^2 \left( 1+	\la^{a}_{n_1} \la^{b}_{n_2}/J^2 \right) \delta_{n_1n_1'} \delta_{n_2'n_2}\delta_{aa'}\delta_{b'b}
-
\phi^{a}_{n_1-n_1'}\phi^{b}_{n_2'-n_2}
\delta_{aa'}\delta_{b'b}
\\
&+
%\frac{t^b}{t^a}
\frac{\la^{b}_{n_2}}{J}
\frac{-\la^{a}_{n_1} \la^{a}_{n_1'}/J^2}{1+\la^{a}_{n_1}\la^{a'}_{n_1'}/J^2}
\sum_{m} \phi^a_{n_1-m} \phi^a_{m-n_1'}
\frac{\lambda^a_m}{J}
\delta_{n_2'n_2}
\delta_{aa'} 
\delta_{b'b}
\\
&+
%\frac{t^a}{t^b}
\frac{\la^{a}_{n_1}}{J}
\frac{-\la^{b}_{n_2'} \la^{b}_{n_2}/J^2}{1+\la^{b}_{n_2'}\la^{b}_{n_2}/J^2}
\sum_{m} \phi^b_{n_2'-m} \phi^b_{m-n_2}
\frac{\lambda^b_m}{J}
\delta_{n_1n_1'}
\delta_{aa'}
\delta_{bb'} 
+
O(\phi^3).
\end{aligned}
\end{align}
By contrast, for the integrable case $h=V$, $M$ is given by
\begin{align}\label{Seq:M-b}
\begin{aligned}
&	J^2 M^{b',a';a,b}_{n_2',n_1';n_1,n_2}
=
 t^2 \left( 1+	\la^{a}_{n_1} \la^{b}_{n_2}/J^2 \right) 
\delta_{n_1n_1'} \delta_{n_2'n_2} 
\delta_{aa'}\delta_{b'b}
\\
&
-t\phi^{a}_{n_1-n_1'}\delta_{n_2n_2'}
\left[ 
1-
\frac{\lambda_{n_2}^{b}}{J}  
\dfrac{-\lambda^{a}_{n_1}\lambda^{a}_{n_1'}/J^2}{1+\lambda^{a}_{n_1}\lambda^{a}_{n_1'}/J^2}\frac{(\lambda^{a}_{n_1}+\lambda^{a}_{n_1'})}{J}
\right] 
\delta_{aa'}\delta_{b'b}
\\
&
- t\phi^{b}_{n_2'-n_2}\delta_{n_1n_1'}
\left[ 1
-
\frac{\lambda_{n_1}^{a}}{J} 
\dfrac{-\lambda^{b}_{n_2}\lambda^{b}_{n_2'}/J^2}{1+\lambda^{b}_{n_2}\lambda^{b}_{n_2'}/J^2}\frac{(\lambda^{b}_{n_2}+\lambda^{b}_{n_2'})}{J}
\right] 
\delta_{aa'}\delta_{b'b}
\\
&
+
\phi^{a}_{n_1-n_1'}\phi^{b}_{n_2'-n_2}
\left[
1
+
\dfrac{-\lambda^{a}_{n_1}\lambda^{a}_{n_1'}/J^2}{1+\lambda^{a}_{n_1}\lambda^{a}_{n_1'}/J^2}
\frac{(\lambda^{a}_{n_1}+\lambda^{a}_{n_1'})}{J}
\dfrac{-\lambda^{b}_{n_2}\lambda^{b}_{n_2'}/J^2}{1+\lambda^{b}_{n_2}\lambda^{b}_{n_2'}/J^2}
\frac{(\lambda^{b}_{n_2}+\lambda^{b}_{n_2'})}{J}
\right] 
\delta_{aa'}\delta_{b'b}
\\
&+
%\frac{t^b}{t^a}
\suml{m}\phi^{a}_{n_1-m}\phi^{a}_{m-n_1'}\delta_{n_2n_2'}
\left[ 
1
+
\frac{ \lambda_{n_2}^{b}}{J} 
\frac{-\la^{a}_{n_1}	\la^{a'}_{n_1'}/J^2 }
{1+\la^{a}_{n_1}	\la^{a'}_{n_1'}/J^2}
\right. 
\\
&\left. \times 
\left\lbrace 
\frac{\lambda^a_m}{J}
+
\left[ 
\dfrac{-\lambda^{a}_{m}\lambda^{a}_{n_1'}/J^2}{1+\lambda^{a}_{m}\lambda^{a}_{n_1'}/J^2} \frac{(\lambda^{a}_{m}+\lambda^{a}_{n_1'})}{J}
+
\dfrac{-\lambda^{a}_{m}\lambda^{a}_{n_1}/J^2}{1+\lambda^{a}_{m}\lambda^{a}_{n_1}/J^2}
\frac{(\lambda^{a}_{m}+\lambda^{a}_{n_1})}{J}
\right] 
-
\frac{\la^{a}_{m}}{J}
\dfrac{(\lambda^{a}_{n_1}+\lambda^{a}_{m})/J}{1+\lambda^{a}_{n_1}\lambda^{a}_{m}/J^2} 
\dfrac{(\lambda^{a}_{m}+\lambda^{a}_{n_1'})/J}{1+\lambda^{a}_{m}\lambda^{a}_{n_1'}/J^2} 
\right\rbrace 
\right] 
\delta_{aa'}\delta_{b'b}
\\
&
+
%\frac{t^a}{t^b}
\suml{m}\phi^{b}_{n_2'-m}\phi^{b}_{m-n_2}\delta_{n_1n_1'}
\left[ 
1
+
\frac{\lambda_{n_1}^{a}}{J}   
\frac{-\la^{b}_{n_2}	\la^{b}_{n_2'}/J^2 }
{1+\la^{b}_{n_2}	\la^{b}_{n_2'}/J^2}
\right. 
\\
&\left. \times 
\left\lbrace 
\frac{\lambda^b_m}{J}
+
\left[ 
\dfrac{-\lambda^{b}_{m}\lambda^{b}_{n_2'}/J^2}{1+\lambda^{b}_{m}\lambda^{b}_{n_2'}/J^2}
\frac{ (\lambda^{b}_{m}+\lambda^{b}_{n_2'})}{J}
+
\dfrac{-\lambda^{b}_{m}\lambda^{b}_{n_2}/J^2}{1+\lambda^{b}_{m}\lambda^{b}_{n_2}/J^2}
\frac{(\lambda^{b}_{m}+\lambda^{b}_{n_2})}{J}
\right] 
-
\frac{\la^{b}_{m}}{J^2}
\dfrac{(\lambda^{b}_{n_2}+\lambda^{b}_{m})/J}{1+\lambda^{b}_{n_2}\lambda^{b}_{m}/J^2} 
\dfrac{(\lambda^{b}_{m}+\lambda^{b}_{n_2'})/J}{1+\lambda^{b}_{m}\lambda^{b}_{n_2'}/J^2} 
\right\rbrace 
\right] 
\delta_{aa'}\delta_{b'b}
+
O(\phi^3).
\end{aligned}
\end{align}

%%%%%%%%%%%%%%%%%
%%%%%%%%%%%%%%%%%
%%%%%%%%%%%%%%%%%

 As explained earlier, the propagator for the Gaussian fluctuation $\D$ is given by the inverse of the kernel $M^{-1}$ (Eq.~\ref{Seq:S2-E}).
 In the non-interacting case (see also the detailed discussion in Refs.~\cite{PRL,arXiv}), after setting $\phi=0$ in Eq.~\ref{Seq:M-a} (or equivalently Eq.~\ref{Seq:M-b}), one finds that the Gaussian fluctuation propagator acquires the form
\begin{align}
\begin{aligned}
	\D^{ab,b'a'}_{n_1n_2,n_2'n_1'}  
	=&
	t^4 (M^{-1})^{ab,b'a'}_{n_1n_2,n_2'n_1'}
	=
	t^2J^2\frac{1}{\left( 1+	\la^{a}_{n_1} \la^{b}_{n_2}/J^2 \right) }
	\delta_{n_1n_1'} \delta_{n_2'n_2}\delta_{aa'}\delta_{b'b}
	\\
	=&
	t^2J^2\frac{1}{\left[ 
		1+s^{a}_{n_1}s^{b}_{n_2}
		+\frac{i}{2J} \left(  s^{b}_{n_2}\zeta_a\e_{n_1}+s^{a}_{n_1}\zeta_b\e_{n_2} \right) 
		\right] }
	\delta_{n_1n_1'} \delta_{n_2'n_2}\delta_{aa'}\delta_{b'b},
\end{aligned}
\end{align}
where in the last equality we have performed a low energy expansion of $\lambda$ (Eq.~\ref{Seq:la}).
It is straightforward to see from the equation above that, in the non-interacting case, the Gaussian fluctuations encoded by $\delta Q^{ab}_{n_1n_2}$ and $\delta Q^{ba}_{n_2n_1}$ are massless as long as $s^{a}_{n_1}=-s^{b}_{n_2}$. By contrast, if $s^{a}_{n_1}=s^{b}_{n_2}$, the associated fluctuation modes are massive, and their contributions to the SFF are negligible. 
For the special saddle point $\Qsp^{(\pm)}$ defined by $s_n^a=\pm \zeta_a$ ($|\e_n|<2J$), all inter-replica ($a\neq b)$ fluctuations $\delta Q^{ab}$ are massless. 
We note that all intra-replica ($a=b$) fluctuations contribute to the disconnected SFF defined as
$K^{\rm dis} (t) \equiv \left\langle \Tr e^{-iHt}\right\rangle\left\langle \Tr e^{iHt}\right\rangle$, and is therefore irrelevant to the connected SFF $K(t)-K^{\rm dis}(t)$ responsible for  the universal ramp.

\subsection{Fluctuations around the saddle point $\Qsp^{(\pm)}$}

We now focus first on the fluctuations $\delta Q$ around the saddle point $\Qsp^{(\pm)}$ ($s_n^a=\pm \zeta_a$ when $|\e_n|<2J$) for both the chaotic and integrable ensembles. 
In the $h\neq V$ chaotic case, after an expansion of Eq.~\ref{Seq:M-a} in powers of Matsubara frequency, we find
\begin{align}\label{Seq:MS1}
\begin{aligned}
M^{b'a',ab}_{n_2'n_1',n_1n_2}
=&
\frac{t^2}{J^2}
 \left[  1+ \zeta_a\zeta_b \pm \frac{i}{2J}\zeta_a\zeta_b \left( \e_{n_1}+\e_{n_2} \right) \right]  
\delta_{n_1n_1'} \delta_{n_2'n_2}\delta_{aa'}\delta_{b'b}
-
\frac{1}{J^2}
\phi^{a}_{n_1-n_1'}\phi^{b}_{n_2'-n_2}
\delta_{aa'}\delta_{b'b}
\\
&-
\frac{1}{2J^2} 
\zeta_a\zeta_b
\sum_{m} \phi^a_{n_1-m} \phi^a_{m-n_1'}
\delta_{n_2'n_2}
\delta_{aa'} 
\delta_{bb'}
-
\frac{1}{2J^2} 
\zeta_a\zeta_b
\sum_{m} \phi^b_{n_2'-m} \phi^b_{m-n_2}
\delta_{n_1n_1'}
\delta_{aa'}
\delta_{bb'} 
+
O(\phi^3,\e\phi,\e^2),
\end{aligned}
\end{align}
whose Fourier transform is 
\begin{align}\label{Seq:MS2}
\begin{aligned}
M^{b'a',ab}_{t_2't_1',t_1t_2}
=&
\frac{1}{J^2}
\left[  
1+ \zeta_a\zeta_b\pm \frac{i}{2J} \zeta_a\zeta_b 
\left(- i\partial_{t_1}+ i\partial_{t_2} \right) 
-
\frac{1}{2}
\zeta_a\zeta_b
\left( \phi^{a}_{t_1}+\zeta_a\zeta_b\phi^{b}_{t_2} \right)^2 
\right]  
\delta_{t_1t_1'} \delta_{t_2't_2}\delta_{aa'}\delta_{bb'}.
\end{aligned}
\end{align}
Similarly for the integrable $h=V$ case, one can see from Eq.~\ref{Seq:M-b} that the kernel $M$ associated with $\Qsp^{(\pm)}$, to the lowest order in energy is given by
\begin{align}\label{Seq:MS3}
\begin{aligned}
&M^{b'a',ab}_{n_2'n_1',n_1n_2}
=
\frac{t^2}{J^2}
 \left[  1+ \zeta_a\zeta_b \pm \frac{i}{2J}\zeta_a\zeta_b \left( \e_{n_1}+\e_{n_2} \right) \right]  
\delta_{n_1n_1'} \delta_{n_2'n_2}\delta_{aa'}\delta_{b'b}
\\
&
+\frac{
1+\zeta_a\zeta_b
}{J^2}
\left[ 
-t\phi^{a}_{n_1-n_1'}\delta_{n_2n_2'}
-t\phi^{b}_{n_2'-n_2}\delta_{n_1n_1'}
+
\phi^{a}_{n_1-n_1'}\phi^{b}_{n_2'-n_2}
+
\suml{m}\phi^{a}_{n_1-m}\phi^{a}_{m-n_1'}\delta_{n_2 n_2'}
+
\suml{m}\phi^{b}_{n_2'-m}\phi^{b}_{m-n_2}\delta_{n_1 n_1'}
\right]
\delta_{aa'}\delta_{b'b}
,
\end{aligned}
\end{align}
and its Fourier transform is
\begin{align}\label{Seq:MS4}
\begin{aligned}
&M^{b'a',a b}_{t_2't_1',t_1t_2}
=
\frac{1}{J^2}
\left\lbrace 1
+\zeta_a\zeta_b\pm\frac{i}{2J} \zeta_a\zeta_b\left( -i\partial_{t_1}+i\partial_{t_2}\right) 
+\left( 
1+\zeta_a\zeta_b
\right) 
\left[ 
-\phi^{a}_{t_1}
-\phi^{b}_{t_2}
+
\phi^{a}_{t_1}\phi^{b}_{t_2}
+
(\phi^{a}_{t_1})^2
+
(\phi^{b}_{t_2})^2
\right] 
\right\rbrace 
\delta_{t_1t_1'} \delta_{t_2't_2}
\delta_{a a'}\delta_{b'b}.
\end{aligned}
\end{align}

Using these results and Eq.~\ref{Seq:DEQ}, we find that the propagator for the inter-replica fluctuation ($\delta \tilde{Q}^{ab}$ with $a\neq b$), i.e., the $0$-dim diffuson propagator $\D^{ab,ba}$, is the solution to the following equation,
\begin{subequations}
\begin{align}
    	&\label{Seq:D-a}
    	\left[  
    	\pm \frac{i}{2J^3}  
    	\left( i\partial_{t_1}- i\partial_{t_2} \right) 
    	+
    	\frac{1}{2J^2}
    	\left( \phi^{a}_{t_1}-\phi^{b}_{t_2} \right)^2 
    	\right]  
    	\mathcal{D}^{ba,a'b'}_{t_2 t_1, t_1' t_2'}
    	=
    	{\bf 1}^{ba,a'b'}_{t_2 t_1, t_1' t_2'},
    	&
    	h\neq V,
    	\\
    	 &\label{Seq:D-b}
     \left[  
    \pm \frac{i}{2J^3} \left( i\partial_{t_1}-i\partial_{t_2}\right) 
    \right] 
    \mathcal{D}^{ba,a'b'}_{t_2 t_1, t_1' t_2'}
    =
    {\bf 1}^{ba,a'b'}_{t_2 t_1, t_1' t_2'},
    &
    h=V.
\end{align}
\end{subequations}
Note that here we have ignored terms of the order $O(\e \phi)$ which contribute to the renormalization effect but is irrelevant to the dephasing.

In the integrable $h=V$ case, $\phi$-dependent term in $M^{ba,ab}$ disappears when $a\neq b$, and the equation for inter-replica fluctuation propagator (Eq.~\ref{Seq:D-b}) is equivalent to that of the non-interacting theory. 
Therefore, the inter-replica fluctuations (diffusons) remain massless in the presence of interactions, indicating the absence of the dephasing effect.

For the chaotic $h\neq V$ case, the $\phi$-dependent term in the diffuson equation (Eq.~\ref{Seq:D-a}) is responsible for the appearance of a mass term for the inter-replica fluctuation, which signals the interaction-induced dephasing effect.
To see this effect, let us now consider the large $t$ limit for simplicity.
We employ a change of variables as in Refs.~\cite{AAK,Kravtsov-DL}:
\begin{align}\label{eq:eta}
\begin{aligned}
	&u=\frac{t_1+t_2}{2},
	\qquad
	\tau=t_1-t_2,
	\qquad
	u'=\frac{t_1'+t_2'}{2},	
	\qquad
	\tau'=t_1'-t_2',	
	\\
	&
	\D^{ba,a'b'}(\tau,\tau';u,u')
    =
	\D^{ba,a'b'}_{t_2t_1,t_1't_2'}.
\end{aligned}
\end{align}	
After the transformation, the diffuson equation for the chaotic $h\neq V$ case (Eq.~\ref{Seq:D-a}) now takes the form 
\begin{align}\label{Seq:DffusonEQ2}
\begin{aligned}
&
\frac{1}{J^3}
\left\lbrace 
\mp
\partial_{\tau} 
+
\frac{J}{2} 
\left[  \phi^a\left(u+\frac{\tau}{2}\right) - \phi^b\left(u-\frac{\tau}{2}\right) \right]^2 
\right\rbrace 
\D^{ba,a'b'}(\tau,\tau';u,u')
=
\delta(u-u')\delta(\tau-\tau')\delta_{aa'}\delta_{bb'}.
\end{aligned}
\end{align}	
Its solution can be expressed as
\begin{align}\label{Seq:D}
\begin{aligned}
	&\D^{ba,a'b'}(\tau,\tau';u,u')
	=
	\exp\left( 
	-S_D[\phi]
	\right)
	\D_0^{ba,a'b'}(\tau,\tau';u,u'),
	\\
	&S_D[\phi]=\mp \frac{J}{2}
	\int_{\tau'}^{\tau} dv
    \left[  \phi^a\left(u+\frac{v}{2}\right) - \phi^b\left(u-\frac{v}{2}\right) \right]^2 .
\end{aligned}
\end{align}
Here $\D_0$ represents the solution to the non-interacting diffuson equation (Eq.~\ref{Seq:D-b}), and is given by
\begin{align}\label{Seq:D0}
    \D_0^{ba,a'b'}(\tau,\tau';u,u')
    =
    \mp s' J^3\Theta\left(s'(\tau-\tau')\right) \delta(u-u')\delta_{aa'}\delta_{bb'}.
\end{align}
We note that the diffuson Eq.~\ref{Seq:DffusonEQ2} is satisfied when $s'$ equals $+1$ and $-1$,
%For the non-interacting theory, $s'$ is determined by the infinitesimal increment ignored in the current paper.
%For the solution to the diffuson equation for interacting chaotic model, i.e., Eq.~\ref{Seq:D}, 
but only one of them ($s'=\mp 1$ corresponding to saddle point $\Qsp^{(\pm)}$) obeys the boundary condition that the diffuson propagator $\D$ is non-divergent in the limit $|\tau-\tau'|\rightarrow \infty$. 
Imposing this boundary condition, we find the solution to the diffuson equation in the presence of an arbitrary bosonic field $\phi$. 
Combining Eqs.~\ref{Seq:D} and~\ref{Seq:D0}, one can see that the exponential factor $\exp(-S_D[\phi])$ decays with increasing $|\tau-\tau'|$ for almost any configuration of the real bosonic field $\phi$ (except for the extreme case where $\phi^{+}=\phi^{-}$ is time independent).

Using the Fourier representation for the H.S. field, $S_D[\phi]$ can be rewritten as
\begin{align}\label{eq:DffusonEQ4}
\begin{aligned}
	&S_D[\phi]
	=
	\mp \frac{J}{2}
	\int_{\tau'}^{\tau} dv
	\left[ 
	\frac{1}{t}
	\suml{m}
	\left(
	\phi^a_m e^{i\ww_m({u+v/2})}-\phi^b_m e^{i\ww_m({u-v/2})}
	\right)
	\right]^2 
	\\
	=&
	\mp \frac{J}{2}
	\frac{1}{t^2}
	\suml{m\neq m'}
	\left(
	e^{i\ww_{m-m'}u}
	\phi^a_m \phi^a_{-m'}
	+
	e^{-i\ww_{m-m'}u}
	\phi^b_{-m} \phi^b_{m'}
	-
	e^{i\ww_{m+m'}u}
	2\phi^a_m \phi^b_{m'}
	\right)
	\frac{e^{i\ww_{m-m'}\tau/2}-e^{i\ww_{m-m'}\tau'/2}}{i\ww_{m-m'}/2}
	\\
	&\mp \frac{J}{2}
	\frac{1}{t^2}
	\suml{m}
	\left(
	\phi^a_m \phi^a_{-m}
	+
	\phi^b_m \phi^b_{-m}
	-
	2\phi^a_m \phi^b_{m}
	e^{i2\ww_{m}u}
	\right)
	(\tau-\tau').
\end{aligned}
\end{align}
At large $|u|$ and $|\tau-\tau'|$, the contribution from the highly oscillating terms can be ignored, and $S_D[\phi]$ is given approximately by the
nonoscillating term:
\begin{align}\label{eq:DffusonEQ5}
\begin{aligned}
&S_D[\phi]
\approx
\mp 
\frac{(\tau-\tau')}{\tau_{\phi}},
\\
&\tau_{\phi}^{-1}= \frac{J}{2}
\frac{1}{t^2}
\left[
\suml{m\neq 0}
\left(
|\phi^a_m|^2
+
|\phi^b_m|^2
\right)
+
\left(
\phi^a_0 
-
\phi^b_0 
\right)^2
\right].
\end{aligned}
\end{align}
As a result, in this case, the diffuson propagator $\D$ acquires a mass term $\tau_{\phi}^{-1}$ and takes the form
\begin{align}
\begin{aligned}
    &\D^{ba,a'b'}(\tau,\tau';u,u')
    \approx
     J^3
    \exp\left(-|\tau-\tau'|/\tau_{\phi}\right)
    \Theta\left(\mp (\tau-\tau')\right) \delta(u-u')\delta_{aa'}\delta_{bb'}.
\end{aligned}
\end{align}

In summary, in the large $t$ limit, the diffuson propagator $\D^{ba,ab}_{u-\tau/2,u+\tau/2,u'+\tau'/2,u'-\tau'/2}$, in the presence of interactions, acquires an exponential factor $e^{-S_D[\phi]}$ which decays with increasing $|\tau-\tau'|$ and can be approximated as $e^{-|\tau-\tau'|/\tau_{\phi}}$ at large $|\tau-\tau'|$ and $|u|$.
The exponential factor $e^{-S_D[\phi]}$ is a manifestation of the dephasing effect which comes from interactions since $\phi$ is the interaction decoupling field.
Moreover, from Eq.~\ref{Seq:IntdQ}, we can see that the exponential factor $e^{-S_D[\phi]}$ results in the suppression of the connected SFF. The calculation of the contribution from the diffusons to the SFF averaged over the fluctuations of the H.S. decoupling field $\phi$ is left to future study.

%%%%%%%%%%%%%%%%%%%%%%%%%%%%%%%%%%%%%%%%%%%%%%%%%

\subsection{Fluctuations around a generic saddle point}

We now consider fluctuations around a generic saddle point, in particular those that are massless in the non-interacting theory, i.e. the 0-dim diffusons.
As explained earlier, these fluctuations are encoded by $\delta Q^{ab}_{n_1n_2}$ if $s_{n_1}^a=-s_{n_2}^b$ is satisfied. 
The relevant kernel elements  $M^{b'a',ab}_{n_2'n_1',n_1n_2}$ associated with these modes are the ones obey the contraint: $s_{n_1}^a=s_{n_1'}^a=-s_{n_2}^b=-s_{n_2'}^b$. 
In this case, $M^{b'a',ab}_{n_2'n_1',n_1n_2}$ (Eqs.~\ref{Seq:M-a} and~\ref{Seq:M-b}) can be simplified as the following after an expansion in terms of energy:
	\begin{align}
	\begin{aligned}
	J^2 M^{b'a',ab}_{n_2'n_1',n_1n_2}
	=&
	\left\lbrace
	t^2
	\left[ -\frac{i}{2J} s_{n_1}^a
	\left( \zeta_a\e_{n_1}-\zeta_b\e_{n_2}  \right)\right] 
	\delta_{n_1n_1'} \delta_{n_2'n_2}
	-
	\phi^{a}_{n_1-n_1'}\phi^{b}_{n_2'-n_2}
    \right.
	\\
	&\left.
	-
	\frac{1}{2} 
	s^{b}_{n_2}
	\sum_{m} \phi^a_{n_1-m} \phi^a_{m-n_1'}
	s^a_m
	\delta_{n_2'n_2}
	-
	\frac{1}{2} 
	s^{a}_{n_1}
	\sum_{m} \phi^b_{n_2'-m} \phi^b_{m-n_2}
	s^b_m
	\delta_{n_1n_1'}
	\right\rbrace
	\delta_{aa'}\delta_{b'b},
	&
	h\neq V,
	\\
	J^2 M^{b'a',ab}_{n_2'n_1',n_1n_2}
	=&
	\left\lbrace
	t^2
	\left[ - \frac{i}{2J} s^{a}_{n_1}
	\left( \zeta_a \e_{n_1}- \zeta_b\e_{n_2}\right) \right] 
	\delta_{n_1n_1'} \delta_{n_2'n_2}
    \right.
	\\
	&\left.
	-
	\suml{s_{n_1}^a s_m^a<0}
	\phi^{a}_{n_1-m}\phi^{a}_{m-n_1'}
	\dfrac{ 2\e_m^2}{\ww_{n_1-m} \ww_{m-n_1'}} 
	\delta_{n_2n_2'}
	-
	\suml{s_{n_2}^b s_m^b<0}
	\phi^{b}_{n_2'-m}\phi^{b}_{m-n_2}
	\dfrac{ 2\e_m^2}{\ww_{n_2'-m}\ww_{m-n_2}} 
	\delta_{n_1n_1'}
	\right\rbrace
	\delta_{aa'}\delta_{b'b},
    &
    h=V.
	\end{aligned}
	\end{align}
Inserting these  expressions above into Eq.~\ref{Seq:S2-E}, the fluctuations around an arbitrary saddle point can therefore be studied in an analogous manner as in the case of $\Qsp^{(\pm)}$. However, in the present paper, we focus on the fluctuations around $\Qsp^{(\pm)}$ due to the complicated structure of the kernel $M$ for other saddle points.

\section{Diagrammatic representation of saddle point and diffuson}\label{Ssec:diagram}

In this section, through diagrammatic calculation, we show that the saddle point $\tilde{Q}_{sp}$ and the Gaussian fluctuation propagator $\D\equiv\left\langle \tilde{\delta Q} \tilde{\delta Q} \right\rangle$ are related to the fermionic self-energy and the diffuson propagator defined by infinite series of impurity ladder diagrams, respectively.

\begin{figure}[h!]
    \centering
	\includegraphics[width=0.72\linewidth]{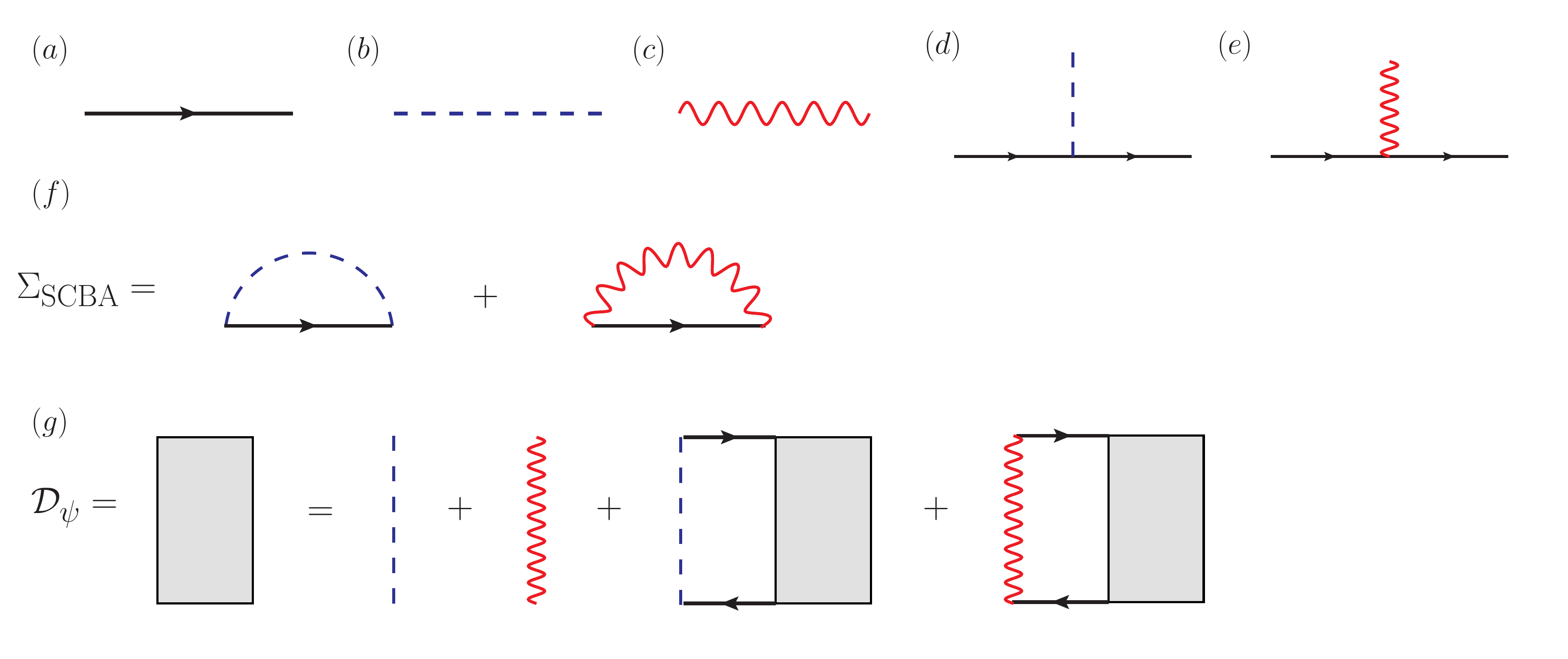}
	\caption{Diagrammatic representations of saddle point $\tilde{Q}_{sp}$ and fluctuation propagator $\D=\left\langle \delta \tilde{ Q} \delta \tilde{ Q} \right\rangle$ for the $h\neq V$ chaotic ensemble. The bold black solid line in panel (a) represents the ensemble averaged fermionic Green's function, while the blue dashed and red wavy lines in panels (b) and (c) show the correlation functions of $h$ and $V$ matrix elements, respectively.
	The vertices coupling fermionic field $\psi$ and random matrix $h$ ($V$) are represented by the diagram in panel (d) [(e)]. 
	The saddle point $\tilde{Q}_{sp}$ is associated with the self-consistent self-energy $\Sigma_{\rm SCBA}$ defined diagrammatically in panel (f). 
	The fluctuation propagator $\D$ is related to the diffuson propagator $\D_{\psi}$ (gray box) defined by the diagrammatic Bethe-Salpeter equation in panel (g). }
	\label{fig:s1}
\end{figure}

For the $h\neq V$ chaotic case, saddle point $\tilde{Q}_{sp}$ and the Gaussian fluctuation propagator $\D$ are defined by the diagrams in Figs.~\ref{fig:s1}(f) and~(g), respectively. The relevant Feynman rules are presented in Fig.~\ref{fig:s1} panels (a)-(e). In particular, the black bold solid line in Fig.~\ref{fig:s1}(a) represents the ensemble averaged fermionic Green's function:
\begin{align}\label{Seq:G2}
\begin{aligned}
	(G_{\psi})^{ab}_{t_1t_2}
	\equiv
	\left\langle \psi_i^a(t_1) \bar{\psi}_i^b(t_2) \right\rangle .
%	=
%	\left(\frac{1}{\partial_t-\Sigma_{\rm SCBA}}\right)_{t_1,t_2}^{ab} 
%	=
%	(\frac{\sigma^3}{\partial_t\sigma^3+F\Qsp})_{t_1,t_2}^{ab} 
\end{aligned}
\end{align}
The blue dashed and red wavy lines in Figs.~\ref{fig:s1}(b) and~(c) correspond to the correlations of matrix elements of $h$ and $V$, respectively,
\begin{subequations}\label{Seq:hh}
\begin{align}
    &\label{Seq:hh-a}
	\left\langle h_{ij} h_{j'i'} \right\rangle 
	=J^2/N \delta_{ii'}\delta_{jj'},
    \\
    &\label{Seq:hh-b}
    \left\langle V_{ij} V_{j'i'} \right\rangle
    =J^2/N \delta_{ii'}\delta_{jj'}.
\end{align}
\end{subequations}
For the vertices depicted in Figs.~\ref{fig:s1}(d) and~(e) which couple $\psi$ and $h/V$, their amplitudes are given by, respectively
\begin{subequations}\label{Seq:vertex}
\begin{align}
	&\label{Seq:vertex-d}
	(d)=i \bar{\psi}_i^a(t') \zeta_a h_{ij} \psi_j^a(t'),
	\\
	&\label{Seq:vertex-e}
	(e)= i \bar{\psi}_i^a(t') \zeta_a  \phi^a(t') V_{ij} \psi_j^a (t').
\end{align}
\end{subequations}
The self-energy for the fermionic field $\psi$ is given by a series of rainbow diagrams, and is defined self-consistently by the diagram in Fig.~\ref{fig:s1}(f) where the black bold solid lines represent the self-energy renormalized fermionic Green's function $G_{\psi}$(Eq.~\ref{Seq:G2}). 
Self-energy diagrams with crossed `impurity' ($h$ or $V$) lines are of higher order in the large $N$ expansion and can therefore be ignored.
One can check that the self-consistent Born approximated (SCBA) self-energy depicted in Fig.~\ref{fig:s1}(f) is given by the following equation:
\begin{align}
\begin{aligned}
    (\Sigma_{\rm SCBA})^{ab}_{t_1t_2}
    =&
%	-(\sigma^3F\Qsp)^{ab}_{t_1t_2}
%	=
	-\suml{j}
	\zeta_a\zeta_b
	\left[ 
	\left\langle h_{ij} h_{ji} \right\rangle 
	+\phi^a(t_1)\phi^b(t_2)\left\langle V_{ij} V_{ji} \right\rangle
	\right] 
	(G_{\psi})^{ab}_{t_1t_2}
	=
	-J^2 \zeta_a\zeta_b
	F^{ab,ba}_{t_1 t_2,t_2 t_1} 
	(G_{\psi})_{t_1t_2}^{ab}, 
\end{aligned}
\end{align}
which can be simplified as 
\begin{align}\label{Seq:SCBA}
\begin{aligned}
    \Sigma_{\rm SCBA}
	=
	-J^2 \left(F \sigma^3 G_{\psi} \sigma^3  \right).
\end{aligned}
\end{align}
Using the Dyson equation 
\begin{align}
    G_{\psi}=\left(\partial_t-\Sigma_{\rm SCBA}\right)^{-1},
\end{align}
 we obtain the self-consistent equation for the self-energy
\begin{align}
\begin{aligned}
     \Sigma_{\rm SCBA}
    =
	-J^2
	 F 
	 \sigma^3
	\left(\partial_t-\Sigma_{\rm SCBA} \right)^{-1}
	\sigma^3
	.
\end{aligned}
\end{align}
Comparing this equation with the saddle point equation for $\Qsp$ (Eq.~\ref{Seq:SPEQ}), we can see that 
\begin{align}
    \Sigma_{\rm SCBA}
    =
	-\sigma^3F\Qsp,
	\qquad
	G_{\psi}=\left(\partial_t+\sigma^3F\Qsp\right)^{-1}= G\sigma^3,
\end{align}    
where $G$ is defined in Eq.~\ref{Seq:G}.
Similarly, one can show that the diffuson propagator $\D_{\psi}$ defined by the diagrammatic Bethe–Salpeter equation in Fig.~\ref{fig:s1} is the solution to the following equation:
\begin{align}
\begin{aligned}
	(\D_{\psi})^{ab,b'a'}_{t_1t_2,t_2't_1'}
	=&
	-\delta_{aa'}\delta_{bb'}\delta_{t_1t_1'}\delta_{t_2 t_2'}
	\zeta_a\zeta_b
	\left[ 
	\left\langle h_{ii'} h_{i'i} \right\rangle 
	+\phi^a(t_1)\phi^b(t_2)\left\langle V_{ii'} V_{i'i} \right\rangle
	\right]
	\\
	&-
	\sum_{a''b''}\int_{t_1'',t_2''}
	\zeta_a\zeta_b
	(G_{\psi})^{aa''}_{t_1t_1''} (G_{\psi})^{b''b}_{t_2''t_2}
	(\D_{\psi})^{a''b'',b'a'}_{t_1''t_2'',t_2't_1'}
	\suml{j}
	\left[ 
	\left\langle h_{ij} h_{ji} \right\rangle 
	+\phi^a(t_1)\phi^b(t_2)\left\langle V_{ij} V_{ji} \right\rangle
	\right],
%	\\
%	\zeta_{a}(\D_{\psi})^{ab,b'a'}_{t_1t_2,t_2't_1'}\zeta_{b'}
%	=&
%	-\delta_{aa'}\delta_{bb'}\delta_{t_1t_1'}\delta_{t_2 t_2'}
%	\left[ 
%	1
%	+\phi^a(t_1)\phi^b(t_2)
%	\right]J^2/N
%	\\
%	&-
%	(G_{\psi})^{aa''}_{t_1t_1''}\zeta_{a''} (G_{\psi})^{b''b}_{t_2''t_2}\zeta_{b}
%	\zeta_{a''}(\D_{\psi})^{a''b'',b'a'}_{t_1''t_2'',t_2't_1'}\zeta_{b'}
%	\left[ 
%	1+\phi^a(t_1)\phi^b(t_2)
%	\right]J^2
\end{aligned}
\end{align}
which can be rewritten as
\begin{align}\label{Seq:BS}
\begin{aligned}
	\sum_{a''b''}\int_{t_1'',t_2''}
    \left[ 
    \frac{1}{J^2}
    1^{ab,b''a''}_{t_1 t_2, t_2'' t_1''}
    +
    F^{ab,ba}_{t_1 t_2,t_2t_1}
    ( G_{\psi}\sigma^3)^{aa''}_{t_1t_1''} (G_{\psi}\sigma^3)^{b''b}_{t_2''t_2}
    \right]
	\left(\zeta_{a''}(\D_{\psi})^{a''b'',b'a'}_{t_1''t_2'',t_2't_1'}\zeta_{b'}\right)
	=&
	-\frac{1}{N}
    F^{ab,b'a'}_{t_1 t_2, t_2' t_1'}.
\end{aligned}
\end{align}
When this equation is compared with the ones for Gaussian fluctuation propagator $\D$ (Eqs.~\ref{Seq:DEQ} and~\ref{Seq:M}), one can immediately see that
\begin{align}
\begin{aligned}
    	D^{ab,b'a'}_{t_1t_2,t_2't_1'}
    	=
    	-N\left(\zeta_{a}(\D_{\psi})^{ab,b'a'}_{t_1t_2,t_2't_1'}\zeta_{b'}\right).
\end{aligned}
\end{align}

\begin{figure}[h!]
    \centering
	\includegraphics[width=0.48\linewidth]{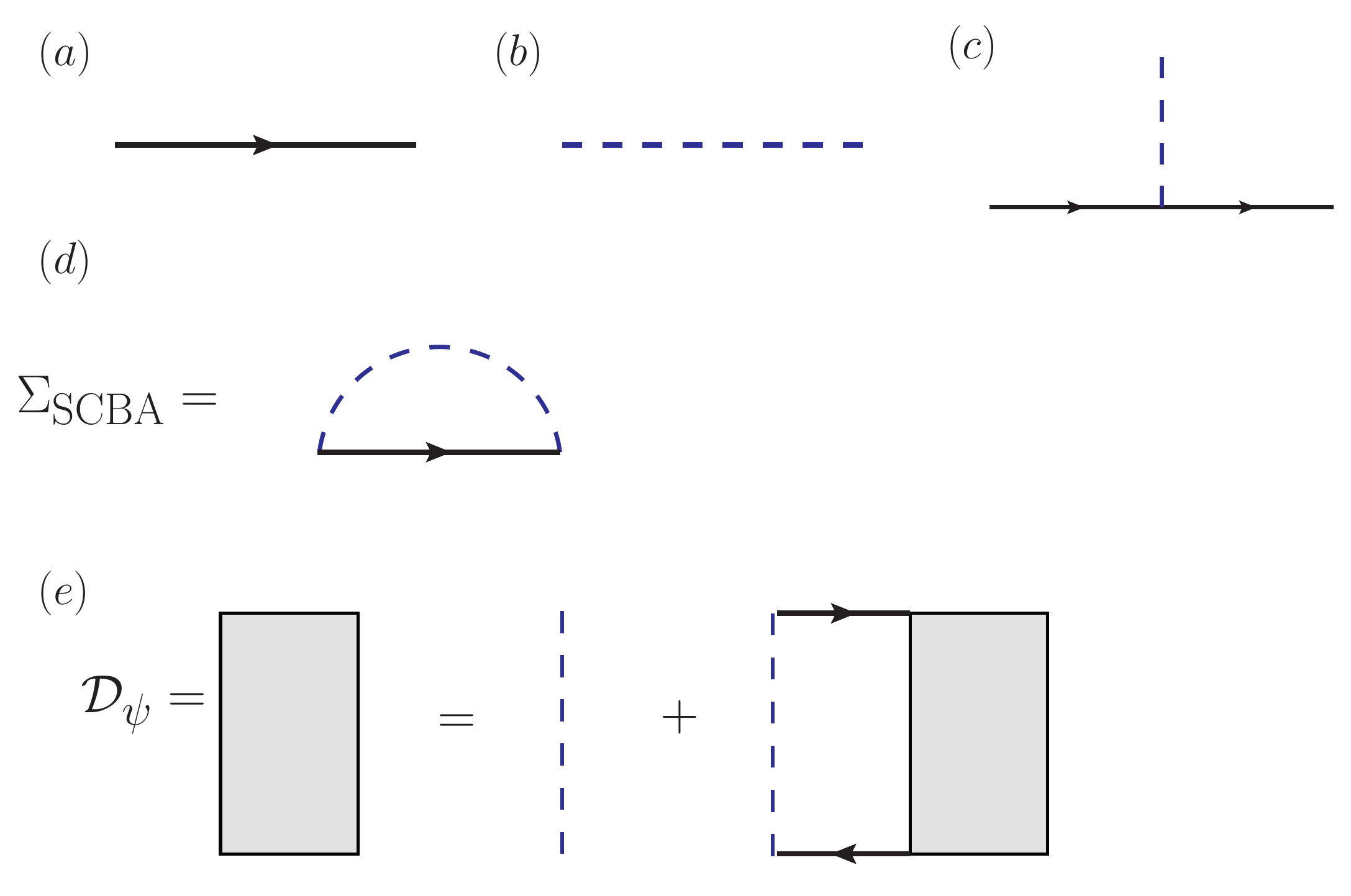}
	\caption{The same as Fig.~\ref{fig:s1} but for the $h=V$ integrable case. In this case $h=V$ is represented by the blue dashed line.}
	\label{fig:s2}
\end{figure}

For the integrable $h=V$ case, one can prove that the equations for the SCBA self-energy and diffuson propagator take the same forms as in the chaotic $h\neq V$ case, just with different expressions for $F$. In this case, we employ the Feynman rules shown in Figs.~\ref{fig:s2} (a)-(c). The bold solid black lines in Fig.~\ref{fig:s2} (a) represent the ensemble averaged fermionic Green's function (Eq.~\ref{Seq:G2}), while the blue dashed line in Fig.~\ref{fig:s2}(b) represents the correlation function of $h=V$ matrix elements (Eq.~\ref{Seq:hh-a}). The vertex depicted in Fig.~\ref{fig:s2} (c) has the amplitude of 
\begin{align}
\begin{aligned}
	(e)= i \bar{\psi}_i^a(t') \zeta_a \left[ 1+\phi^a(t')\right] h_{ij} \psi_j^a (t').
\end{aligned}
\end{align}
Note the difference between this equation and Eq.~\ref{Seq:vertex}.
In an analogous way as in the case of $h\neq V$, one can find that the SCBA self-energy defined in Fig.~\ref{fig:s2}(d) is given by the following equation
\begin{align}
\begin{aligned}
    (\Sigma_{\rm SCBA})^{ab}_{t_1t_2}
    =&
%	-(\sigma^3F\Qsp)^{ab}_{t_1t_2}
%	=
	-\suml{j}
	\zeta_a\zeta_b
	\left[ 
	\left(1+\phi^a(t_1)\right)\left(1+\phi^b(t_2)\right)
	\right] 
	\left\langle h_{ij} h_{ji} \right\rangle 
	(G_{\psi})^{ab}_{t_1t_2},
\end{aligned}
\end{align}
which can also be expressed as Eq.~\ref{Seq:SCBA} as in the chaotic case but with different expression for $F$ (Eqs.~\ref{Seq:f} and~\ref{Seq:F}).
Similarly, the Bethe–Salpeter equation depicted diagrammatically in Fig.~\ref{fig:s2}(e) is 
\begin{align}
\begin{aligned}
	\D^{ab,b'a'}_{t_1t_2,t_2't_1'}
	=&
	-\delta_{aa'}\delta_{bb'}\delta_{t_1t_1'}\delta_{t_2 t_2'}
	\zeta_a\zeta_b
	\left[ 
	\left(1+\phi^a(t_1)\right)\left(1+\phi^b(t_2)\right) 
	\right]
	\left\langle h_{ii'} h_{i'i} \right\rangle 
	\\
	&-
	\sum_{a''b''}\int_{t_1'',t_2''}
	\zeta_a\zeta_b
	(G_{\psi})^{aa''}_{t_1t_1''} (G_{\psi})^{b''b}_{t_2''t_2}
	(\D_{\psi})^{a''b'',b'a'}_{t_1''t_2'',t_2't_1'}
	\suml{j}
	\left[ 
	\left(1+\phi^a(t_1)\right)\left(1+\phi^b(t_2)\right) 
	\right]
	\left\langle h_{ij} h_{ji} \right\rangle ,
\end{aligned}
\end{align}
which can also be simplified to Eq.~\ref{Seq:BS}.
We have therefore proved that the saddle point $\tilde{Q}_{sp}$ and the Gaussian fluctuation propagator $\D$ are indeed related to the fermionic self-energy $\Sigma_{\rm SCBA}$ and diffuson propagator $\D_{\psi}$, for both the chaotic and integrable models.

\section{Comparing time reparametrization and perturbative approaches for the integrable model}

As mentioned earlier, the SFF of the integrable $h=V$ model can be calculated using the time-reparametrization approach. Therefore, one is able to compare the results from the perturbative approach with that from the time-reparametrization approach. In this section, we show that these two approaches yield consistent results.

In the reparametrized time representation, the action $S[\bar{Q},\phi]$ reduces to that of the non-interacting theory with shifted time $t \rightarrow t+\phi_0^a$, and the corresponding saddle point is given by
\begin{align}
\begin{aligned}
	(\bQsp)_{nm}^{ab}
	=
	\delta_{ab}\delta_{nm}
	\tau^a(t) {\bla}_n^a,
	\qquad
	(\bQsp)_{\tau_1^a \tau_2^b}^{ab}
%	=
%	\frac{1}{\tau^a(t) \tau^b(t)}
%	\sum_{n,m}
%	(\bQsp)_{n,m}^{ab}
%	\exp \left( -i\be_n^a \tau_1^a+i\be_m^b \tau_2^b\right) 
	=
	\delta_{ab}
	\frac{1}{\tau^a(t)}
	\sum_{n} {\bla}_n^a 
	\exp \left( -i\be_n^a (\tau_1^a-\tau_2^a)\right).
\end{aligned}
\end{align}
Here $\tau^a(t')$ is defined in Eq.~\ref{Seq:TR} and its value at $t$ is simply $\tau^a(t)=t+\phi^a_0$. 
$\bar{\e}_n^a$ and $\bar{\lambda}_n^a$ are defined similarly as $\e_n$ and $\lambda_n^a$, but with shifted time, $t \rightarrow t+\phi_0^a$,
\begin{align}
\begin{aligned}
	&
	\be^a_n
%	\equiv \frac{2\pi}{\tau^a(t)} (n+\frac{1}{2})
	=\frac{2\pi}{t+\phi_0^a} (n+\frac{1}{2}),
	\qquad
	\bla_{n}^{a}
	=
	\frac{1}{2}
	\left( 
	i\zeta_a\be_n^a 
	+
	s_n^a
	\sqrt{4J^2 -(\be_n^a)^2}
	\right),
\end{aligned}
\end{align}
where as before $s_n^a$ can take both the values of $+1$ and $-1$ when $|\be_n^a|<2J$.

Transforming back to the original time representation, we have
\begin{align}
\begin{aligned}
	(\Qsp)_{t_1t_2}^{ab}
	=
	(\bQsp)_{\tau^a(t_1),\tau^b(t_2)}^{ab}
%	=
%	(\bQsp)_{\int_0^{t_1} dt' (1+\phi^a(t')),\int_0^{t_2} dt' (1+\phi^b(t'))}^{ab}
	=
	\delta_{ab}
	\frac{1}{\tau^a(t)}
	\sum_{n} \bla_n^a 
	\exp \left( -i\be_n^a \int_{t_2}^{t_1} dt' (1+\phi^a(t'))\right),
\end{aligned}
\end{align}
which after Fourier transform becomes
\begin{align}
\begin{aligned}
	&(\Qsp)_{nn'}^{ab}
	=
	\intl{t_1,t_2} (\Qsp)_{t_1t_2}^{ab}
	\exp \left( i\e_n t_1- i\e_{n'} t_2\right) 
	=
	\intl{t_1,t_2} 
	(\bQsp)_{\tau^a(t_1)\tau^b(t_2)}^{ab}
	\exp \left( i\e_n t_1- i\e_{n'} t_2\right) 
%	\\
%	=
%	\intl{t_1,t_2} 
%	(\bQsp)_{\int_0^{t_1} dt' (1+\phi^a(t')),\int_0^{t_2} dt' (1+\phi^b(t'))}^{ab}
%	\exp \left( i\e_n^a t_1- i\e_{n'}^b t_2\right) 
	\\
	=&
	\intl{t_1,t_2} 
	\delta_{ab}
	\frac{1}{\tau^a(t)}
	\sum_{k}  \bla_k^a 
	\exp \left( -i\be_k^a \int_{t_2}^{t_1} dt' (1+\phi^a(t'))\right)
	\exp \left( i\e_n t_1- i\e_{n'} t_2\right) 
	\\
	=&
	\intl{t_1,t_2} 
	\delta_{ab}
	\frac{1}{\tau^a(t)}
	\sum_{k}  \bla_k^a 
	\exp \left[ -i\be_k^a 
	\left( t_1-t_2 +\suml{m \neq 0}\phi^a_m \frac{e^{-i\ww_m t_1}-e^{-i\ww_m t_2}}{-i\ww_m t}
	+\frac{\phi_0^a}{t}(t_1-t_2)
	\right) \right] 
	\exp \left( i\e_n t_1- i\e_{n'} t_2\right).
\end{aligned}
\end{align}
Here in the last equality, we have employed the Fourier transformation $\phi_{t'}^a=\frac{1}{t}\sum_{m}\phi_m^a e^{-i\ww_m t'}$.
 
 We then expand the equation above in terms of $\phi$ up to order $O(\phi^2)$:
\begin{align}
\begin{aligned}
	(\Qsp)_{nn'}^{ab}
% 	=&
% 	\intl{t_1,t_2} 
% 	\delta_{ab}
% 	\frac{1}{\tau^a(t)}
% 	\sum_{k}  \bla_k^a 
% 	\exp \left[ -i\be_k^a \left( t_1-t_2\right)(1+\frac{\phi_0^a}{t}) \right] 
% 	e^{i\e_n t_1- i\e_{n'} t_2}
% 	\\
% 	\times &
% 	\left\lbrace 
% 	1
% 	-
% 	i\be_k^a \left( \suml{m \neq 0}\phi^a_m 
% 	\frac{e^{-i\ww_m t_1}-e^{-i\ww_m t_2}}{-i\ww_m t} \right) 
% 	-
% 	\frac{1}{2}
% 	(\be_k^a )^2\suml{m,m'\neq0}
% 	\phi^a_m \frac{e^{-i\ww_m t_1}-e^{-i\ww_m t_2}}{-i\ww_m t} \phi^a_{m'}
% 	\frac{e^{-i\ww_{m'} t_1}-e^{-i\ww_{m'} t_2}}{-i\ww_{m'} t} 
% 	+O(\phi^3)
% 	\right\rbrace 
% 	\\
	=\,&
	\intl{t_1,t_2} 
	\delta_{ab}
	\frac{1}{t}\left( 1-\frac{\phi^a_0}{t}+\left( \frac{\phi^a_0}{t}\right) ^2+O(\phi^3)\right) 
	\sum_{k} 
	\left[ 
	\la_k^a
	+
	\frac{d\la_k^a}{d\e_k} \e_k
	\left(-\frac{\phi^a_0}{t}+\left( \frac{\phi^a_0}{t}\right) ^2\right)
	+
	\frac{1}{2}\frac{d^2\la_k^a}{d(\e_k)^2} 
	(\e_k)^2
	\left(\frac{\phi^a_0}{t}\right)^2
	+O(\phi^3)
	\right] 
	\\
	&\times
	\exp \left[ -i
	\e_k\left( 1-\frac{\phi^a_0}{t}+\left( \frac{\phi^a_0}{t}\right)^2+O(\phi^3)\right) 
	 \left( t_1-t_2\right)(1+\frac{\phi_0^a}{t}) \right] 
	e^{i\e_n t_1- i\e_{n'} t_2}
	\\
	\times &
	\left\lbrace 
	\begin{aligned}
	&1
	-
	i\e_k\left( 1-\frac{\phi^a_0}{t}+\left( \frac{\phi^a_0}{t}\right) ^2+O(\phi^3)\right)  \left( \suml{m \neq 0}\phi^a_m \frac{e^{-i\ww_m t_1}-e^{-i\ww_m t_2}}{-i\ww_m t} \right) 
	\\
	&
	-
	\frac{1}{2}
	(\e_k)^2\left( 1-2\frac{\phi^a_0}{t}+3\left( \frac{\phi^a_0}{t}\right)^2+O(\phi^3)\right)\suml{m,m'\neq0}
	\phi^a_m \frac{e^{-i\ww_m t_1}-e^{-i\ww_m t_2}}{-i\ww_m t} \phi^a_{m'}
	\frac{e^{-i\ww_{m'} t_1}-e^{-i\ww_{m'} t_2}}{-i\ww_{m'} t} 
	+O(\phi^3)
	\end{aligned}
	\right\rbrace,
\end{aligned}
\end{align}
where we have used
\begin{align}\label{eq:barlambda}
\begin{aligned}
	&\frac{1}{\tau^a(t)}
	=\frac{1}{t+\phi^a_0}
	\approx
	\frac{1}{t}\left( 1-\frac{\phi^a_0}{t}+\left( \frac{\phi^a_0}{t}\right)^2+O(\phi^3)\right) ,
	\\
	&
	\be_n^a
	%=\frac{2\pi (n+1/2)}{\tau^a(t)}
	=\frac{2\pi (n+1/2)}{t+\phi^a_0}
	=
	\e_n \left( 1-\frac{\phi^a_0}{t}+\left( \frac{\phi^a_0}{t}\right)^2+O(\phi^3)\right) ,
	\\
	&
	\bla_n^a=
	\la_n^a
	+
	\frac{d\la_n^a}{d\e_n} \e_n
	\left(-\frac{\phi^a_0}{t}+\left( \frac{\phi^a_0}{t}\right)^2\right)
	+
	 \frac{1}{2}\frac{d^2\la_n^a}{d(\e_n)^2} 
	 (\e_n)^2
	 \left(\frac{\phi^a_0}{t}\right)^2
	 +O(\phi^3).
\end{aligned}	
\end{align}

Examining terms at each order in $\phi$ separately, we find
	\begin{align}\label{Seq:SP-B}
	\begin{aligned}
	 (Q_{sp}^{(0)})_{nn'}^{aa'}
	 =&
	 \delta_{nn'}\delta_{aa'}\lambda_{n}^{a}t,
	\\
	 (Q_{sp}^{(1)})^{aa'}_{nn'}
	=&
	\delta_{aa'}
	\phi^a_{n-n'}
	\frac{	 
		\lambda_{n'}^a \e_{n'}
		-	
		\lambda_{n}^a \e_{n}
	}
	{\ww_{n-n'}} 
	(1-\delta_{nn'})
	-
	\delta_{aa'}
	\phi^a_0
	\left( 
	\lambda_{n}^a 
	+
	\frac{d\la_n^a}{d\e_n} \e_n
	\right) 
	\delta_{nn'},
	\\
	(Q_{sp}^{(2)})^{aa'}_{nn'}
	=&
	\delta_{aa'}
	\frac{1}{2t}
	\sum_{m\neq n,n'}  
	\frac{\phi^a_{n-m}\phi^a_{m-n'}}{\ww_{n-m} \ww_{m-n'}}
	\left[ 
	\la_{n}^a (\e_{n} )^2
	+
	\la_{n'}^a (\e_{n'} )^2
	-
	2\la_{m}^a (\e_{m})^2
	\right] 
	\\
	&-
	\delta_{aa'}
	\frac{\phi_0^a}{t}
	\frac{\phi^a_{n-n'}}{\ww_{n-n'}} 
	\left[ 
	\e_{n'}
	\left( 
	2\la_{n'}^a 
	+
	\frac{d\la_{n'}^a }{d\e_{n'}} \e_{n'}
	\right) 
	-
	\e_{n}
	\left( 
	2\la_{n}^a 
	+
	\frac{d\la_{n}^a }{d\e_{n}} \e_{n}
	\right) 
	\right] 
	(1-\delta_{nn'})
	\\
	&+
	\delta_{aa'}
	t
	\left( \frac{\phi_0^a}{t}\right)^2   
	\left[ 
	\la_n^a 
	+2\frac{d\la_n^a }{d\e_n} \e_n
	+\frac{1}{2} \frac{d^2\la_n^a }{d(\e_n)^2} (\e_n)^2
	\right] 
	\delta_{nn'}.
	\end{aligned}	
	\end{align}
Using the explicit expression for $\lambda_n^a$ (Eq.~\ref{Seq:la}), one can prove that obtained saddle point expression in Eq.~\ref{Seq:SP-B} is equivalent to the one in Eq.~\ref{Seq:QSPb}.

Let us now consider the quadratic fluctuations around the saddle points.
In the reparametrized time representation, a calculation analogous to the one in Sec.~\ref{Ssec:QF} leads to the action for quadratic fluctuation $\delta \bar{Q}=\bar{Q}-\bQsp$
\begin{align}\label{eq:SbQ}
\begin{aligned}
S[\bar{Q},\phi]
=&
S[\bar{Q}_{sp},\phi]
+
\frac{N}{2J^2}
\suml{ab} \int_{ \tau_1^a, \tau_2^b , \tau_1^a\,', \tau_2^b\,'}
\delta \bQ^{a,b}_{\tau_1^a\,',\tau_2^b\,'} 
\left[ 
\delta(\tau_1^a-\tau_1^a\,')\delta(\tau_2^b-\tau_2^b\,')
+
\frac{1}{J^2}(\bQsp)^{aa}_{\tau_1^a,\tau_1^a\,'} (\bQsp)^{b b}_{\tau_2^b\,',\tau_2^b}
\right] 
\delta \bQ^{b,a}_{\tau_2^b,\tau_1^a}.
\end{aligned}
\end{align}

Converting back to the original time representation, we have
\begin{align}\label{eq:SbQ}
\begin{aligned}
\delta S[\delta Q,\phi]
=&
\frac{N}{2J^2}
\suml{ab} \int_0^{t} dt_1 (1+\phi^a_{t_1})\int_0^{t} dt_2 (1+\phi^b_{t_2}) \int_0^{t} dt_1' (1+\phi^a_{t_1'}) \int_0^{t}dt_2'(1+\phi^b_{t_2'}) 
\\
&\times
\delta Q^{a,b}_{t_1',t_2'} 
\left[ 
\dfrac{\delta(t_1-t_1')}{1+\phi^a_{t_1}}
\dfrac{\delta(t_2-t_2')}{1+\phi^b_{t_2}}
+
\frac{1}{J^2}(\bQsp)^{aa}_{\tau^a(t_1),\tau^a(t_1')} (\bQsp)^{b b}_{\tau^b(t_2'),\tau^b(t_2)}
\right] 
\delta Q^{b,a}_{t_2,t_1}
\\
=&
\frac{N}{2}
\suml{ab} \int_{t_1,t_2,t_1',t_2'}
\delta \tQ^{a,b}_{t_1',t_2'} 
M^{b,a;a,b}_{t_2',t_1';t_1,t_2}
\delta \tQ^{b,a}_{t_2,t_1},
\end{aligned}
\end{align}
where 
\begin{align}\label{Seq:M-B}
\begin{aligned}
	M^{b,a;a,b}_{t_2',t_1';t_1,t_2}
	=
	\frac{1}{J^2}
	\left[
	\dfrac{\delta(t_1-t_1')}{1+\phi^a_{t_1}}
	\dfrac{\delta(t_2-t_2')}{1+\phi^b_{t_2}}
	+
	\frac{1}{J^2}(\bQsp)^{aa}_{\tau^a(t_1),\tau^a(t_1')} (\bQsp)^{b b}_{\tau^b(t_2'),\tau^b(t_2)}
	\right].
\end{aligned}
\end{align}
Earlier in this section, we have showed that, in an expansion up to quadratic order in $\phi$, the saddle point solution from
$
	(\Qsp)_{t_1,t_2}^{ab}
	=
	(\bQsp)_{\tau^a(t_1),\tau^b(t_2)}^{ab}
$
is consistent with the previous perturbative result, for the integrable $h=V$ model. Therefore, one can immediate see that $M$ given by Eq.~\ref{Seq:M-B} is equivalent to the perturbative result Eq.~\ref{Seq:M-b} up to order $O(\phi^2)$.

\section{${\floor{N/2}}$ order polynomial ramp in the non-interacting SFF}\label{sec:nonint}

In Ref.~\cite{PRL}, we have obtained the explicit expression for the SFF of the non-interacting model in the large $N\rightarrow \infty$ limit. In this section, we review that derivation and prove that the ramp in the non-interacting SFF is given by a polynomial in $t$ at order $N/2$ and $(N-1)/2$ for even and odd $N$, respectively. 

In the non-interacting theory, one can show that the SFF is given by~\cite{PRL}
\begin{align}\label{eq:FT-0}
\begin{aligned}
K (t)
=\,&
2^N
\bigg\langle
\prod_{i=1}^{N}
\left[ 1+ \cos \left(\e_i t \right) \right]
\bigg\rangle,
\end{aligned}
\end{align}
where $\e_i$ represents the eigenvalue of the GUE matrix $h$, i.e., the single-particle energy level.
In terms of the correlation functions of the single-particle energy level, $K(t)$ can be rewritten as
\begin{align}\label{Seq:Kt}
\begin{aligned}
	K (t)
	=2^N \left[  1+\sum_{n=1}^{N} \frac{1}{n!} \bar{\rr}_n\right] 
	=2^N \exp \left[ \sum_{n=1}^{N} (-1)^{n-1} \frac{1}{n!}\bar{\tm}_n \right],
\end{aligned}
\end{align}
where 
\begin{align}\label{eq:rtn}
\begin{aligned}
	&\bar{\rr}_n
	=	\int  d \e_1 d \e_2 ...d \e_n
	\R_n(\e_1,\e_2,...,\e_n)
	\prod_{i=1}^{n} \cos\left( \e_i t  \right)  ,
	\\
	&\bar{\tm}_n
	=	\int  d \e_1 d \e_2 ...d \e_n
	\Tm_n(\e_1,\e_2,...,\e_n)
	\prod_{i=1}^{n} \cos\left( \e_i t  \right) .
\end{aligned}
\end{align}
Here $\R_n(\e_1,...,\e_n)$ is the $n$-point correlation function of the single-particle energy levels, which can be obtained by integrating the joint probability density function of all single-particle energy levels $P(\e_1, ..., \e_N)$ over $N-n$ arguments
\begin{align}
\R_n(\e_1, ....,\e_n)
=\,
\frac{N!}{(N-n)!}\int d\e_{n+1} ...d\e_N  P(\e_1, ..., \e_N).
\end{align}
$\Tm_n(\e_1,...,\e_n)$ is the $n$-point cluster function of single-particle energy levels, i.e., the connected part of $\R_n$.
Note that $\R_n$ and $\Tm_n$  are defined differently from $R_n$ (Eq.~\ref{Seq:Rn}) and $R_n^{\msf{con}}$ which appear in Eqs.~\ref{Seq:CumSum-0} and~\ref{Seq:CumSum}.
In the second equality of Eq.~\ref{Seq:Kt}, it has been used that the upper limit of the summation in the first equality is $N\rightarrow \infty$ (see Ref.~\cite{PRL,Mehta}).

For the GUE matrix $h$, the explicit expression for the $n$-point single-particle level correlation function can be expressed as
\begin{align}\label{eq:Rn}
\begin{aligned}
\R_n(\e_1,\e_2,...,\e_n)
=\,
\det \left[  \kk(\e_i,\e_j) \right]_{i,j=1,...n},
\end{aligned}
\end{align}
where the matrix elements of $\kk$ are given by, in the large $N$ limit, 
\begin{align}\label{eq:KK}
\begin{aligned}
	\kk(\e_i,\e_j)
	=
	\begin{cases}
	\R_1(\e_i)
	=
	\dfrac{N}{2\pi} \sqrt{4-\e_i^2} \, \Theta (2-|\e_i|),
	&
	i=j,
	\\
	\\
	\kk(\e_i-\e_j)=\dfrac{N}{\pi}\dfrac{\sin \left[ N\left(\e_i-\e_j \right)  \right] }{N \left(\e_i-\e_j \right)},
	&
	i \neq j.
\end{cases}
\end{aligned}
\end{align}
Similarly, the $n$-point cluster function can be expressed as,
\begin{align}\label{eq:Tn}
\begin{aligned}
\Tm_n(\e_1,...\e_n)
=\,
\suml{\PP(n)} \kk(\e_{1},\e_{2})\kk(\e_{2},\e_{3})...\kk(\e_{n-1},\e_{n})\kk(\e_{n},\e_{1}),
\end{aligned}
\end{align}
where the summation runs over all cyclic permutations $\PP(n)$ of the indices $\left\lbrace 1,2,...n \right\rbrace$.

In the intermediate ramp regime, using Eqs.~\ref{eq:Rn} and~\ref{eq:KK}, and writing out the determinant explicitly, one can prove that $\bar{\rr}_n$ (Eq.~\ref{eq:rtn}) is approximately a polynomial in $t$ at order $\floor{n/2}$. More specifically, for even $n$, the leading order in $t$ contribution to $\bar{\rr}_n$ is from terms of the form
\begin{align}
    \prod_{i=1}^{n/2}
    \left[
    \int d\e_{2i-1} d\e_{2i}
    \kk^2(\e_{2i-1},\e_{2i})
    \cos (\e_{2i-1} t)
    \cos (\e_{2i} t)
    \right]
    \propto t^{n/2}.
\end{align}
Here each double integral over the pair $\e_{2i-1}, \e_{2i}$ gives rise to a factor $t$, and their product is of the order of $t^{n/2}$.
Note that we consider the time regime where the Fourier transform of $\R_1(\e)$ can be ignored and the Fourier transform of $\kk^2(\e_{2i-1},\e_{2i})$ is approximately linear in $t$.
Similarly, for odd integer $n$, the leading contribution is from
\begin{align}
    \left[
    \int_{\e_{1},\e_{2},\e_3}
    \kk(\e_{1},\e_{2})
    \kk(\e_{2},\e_{3})
    \kk(\e_{3},\e_{1})
    \cos (\e_{1} t)
    \cos (\e_{2} t)
    \cos (\e_{3} t)
    \right]
    \prod_{i=2}^{(n-1)/2}
    \left[
    \int_{\e_{2i},\e_{2i+1}}
    \kk^2(\e_{2i},\e_{2i+1})
    \cos (\e_{2i} t)
    \cos (\e_{2i+1} t)
    \right]
    \propto t^{(n-1)/2}.
\end{align}
Now the integration over $\e_1$, $\e_2$ and $\e_3$ contributes a factor of $t$, and each of the remaining double integrations  yields a factor of $t$ as well, leading to a contribution of the order of $t^{(n-1)/2}$.
Therefore, one can draw the conclusion that the ramp of the non-interacting SFF is given by a polynomial in $t$ at order $\floor{N/2}$, which is consistent with the periodic-orbit estimation, as explained in the main text.
Note that, in Ref.~\cite{PRL}, the second equality in Eq.~\ref{Seq:Kt} is employed to compute the SFF. It is proved there, with the help of Eq.~\ref{eq:Tn}, that $\bar{\tm}_n$ is approximately linear in $t$ in the intermediate ramp regime. As a result, we obtained a exponential in $t$ function which is an approximation to the $\floor{N/2}$ order polynomial in the large $N$ limit.
\bibliography{main}